\documentclass[12pt,twoside]{article}
\usepackage{correl}

\usepackage{braket}
\usepackage{xcolor}
\usepackage{grffile}
\usepackage{subcaption}
\usepackage{hyperref}

\usepackage{mathtools}
\mathtoolsset{showonlyrefs}

\DeclareMathOperator{\spn}{span}
\DeclareMathOperator{\aut}{Aut}
\DeclareMathOperator{\tr}{Tr}

\graphicspath{{figures/}}

\def\TheTitle{Studying Continuous Symmetry Breaking using Energy Level Spectroscopy}

\def\TheAuthor{A.~Wietek, M.~Schuler, and A.~M.~L\"auchli}
\def\TheInstitute{Institut f\"ur Theoretische Physik}
\def\TheAddress{Universit\"at Innsbruck}
\def\TheAbstract{
\noindent Tower of States analysis is a powerful tool for investigating
  phase transitions in condensed matter systems. Spontaneous symmetry breaking
  implies a specific structure of the energy eigenvalues and their corresponding
  quantum numbers on finite systems. In these lecture notes we explain the
  group representation theory used to derive the spectral structure for
  several scenarios of symmetry breaking. We give numerous examples to compute
  quantum numbers of the degenerate groundstates, including translational symmetry
  breaking or spin rotational symmetry breaking in Heisenberg antiferromagnets.
  These results are then compared to actual numerical data from Exact
  Diagonalization.
}
\def\TheReference{Lecture Notes of the Autumn School on Correlated Electrons
2016, ``Quantum Materials: Experiments and Theory'', Forschungszentrum J\"ulich
(2016)}

\begin{document}
\MakeTitle           

\section{Introduction} \index{tower of states}
Spontaneous symmetry breaking is amongst the most important \index{spontaneous
symmetry breaking}
and fundamental concepts in condensed matter physics. The fact that a
ground- or thermal state
of a system does not obey its full symmetry explains
most of the well-known phase transitions in solid state physics like
crystallization of a fluid, superfluidity, magnetism, superconductivity
and many more. A standard concept for investigating spontaneous
 symmetry breaking is the notion of an order parameter. In the 
thermodynamic limit it is non-zero in the symmetry-broken 
phase and zero in the disordered phase. 

Another concept to detect spontaneous symmetry breaking less widely known but
equally powerful is the \textit{tower of states analysis}
(TOS)~\cite{Anderson1952,Lhuillier2005}.  The energy spectrum, i.e.~the
eigenvalues of the Hamiltonian, \index{energy spectrum} of a {\em finite}
system has a characteristic and systematic structure in a symmetry broken
phase: several eigenstates are quasi-degenerate on finite systems, become
degenerate in the thermodynamic limit and possess certain quantum numbers.  The
TOS analysis deals with understanding the spectral structure of the Hamiltonian
and predicting quantum numbers of the groundstate manifold.  Also on finite
systems spontaneous symmetry breaking manifests itself in the structure of the
energy spectra which are accessible via numerical simulations. Most prominently,
the Exact Diagonalization method \cite{Laeuchli2011,Sandvik2010}
\index{exact diagonalization}
\index{numerical simulations}
can exactly calculate these
spectra, including their quantum numbers, on moderate system sizes.
Different well-established numerical techniques like the Quantum
Monte Carlo technique also allow for performing energy level spectroscopy
to a certain extent~\cite{Suwa2015}.
The predictions of TOS analysis are highly nontrivial statements
which can be used to unambiguously identify symmetry broken 
phases. Thus TOS analysis is a powerful technique to investigate
many condensed matter systems using numerical simulations.
The goal of these lecture notes is to explain the specific
structure of energy spectra and their quantum numbers 
in symmetry broken phases. The anticipated structure
is then compared to several actual numerical simulations using 
Exact Diagonalization. 

These lecture notes have been written at the kind request of the organizers
of the J\"ulich 2016 "Autumn School on Correlated Electrons"~\cite{Pavarini2016}. The notes build 
on and complement  previously available lecture notes by 
Claire Lhuillier~\cite{Lhuillier2005}, by Gr\'egoire  Misguich and Philippe 
Sindzingre~\cite{Misguich2007} and by Karlo Penc and one of the 
authors~\cite{Penc2011}.

The outline of these notes is as follows: in Section~\ref{sec:TOS} we introduce the tower of states of
continuous symmetry breaking and derive its scaling behaviour.
We investigate a toy model which shows most of the relevant features.
Section~\ref{sec:sym_analysis} explains in detail how the multiplicities
and quantum numbers in the TOS can be predicted by elementary
group theoretical methods. To apply these methods we discuss
several examples in Section~\ref{sec:examples} and compare them
to actual numerical data from Exact Diagonalization.


\section{Tower of states} \label{sec:TOS} \index{tower of states!scaling}
We start our discussion on spontaneous symmetry breaking by
investigating the Heisenberg model on the square
\index{Heisenberg model}
lattice. Its Hamiltonian is given by 
\begin{equation}
  \label{eq:heisenberg}
  H = J\sum\limits_{\left< i, j \right>} \mathbf{S}_i \cdot \mathbf{S}_j
\end{equation}
and is invariant under global SU($2$) spin rotations, i.e. a rotation
of every spin on each site with the same rotational SU($2$) matrix.
Therefore the total spin 
\begin{equation}
  \label{eq:totalspin}
  \mathbf{S}_{\text{tot}}^2 = \left(\sum\limits_{i} \mathbf{S}_i\right)^2=
S_{\text{tot}}(S_{\text{tot}}+1)
\end{equation}
is a conserved quantity of this model and every state in the
spectrum of this Hamiltonian can be labeled via its total spin quantum 
number $S_{\text{tot}}$. The Heisenberg Hamiltonian on the square
lattice has the property of being \textit{bipartite}: The lattice
can be divided into two sublattices $A$ and $B$ such that every term in 
Eq.~\eqref{eq:heisenberg} connects one site from sublattice $A$ to 
sublattice $B$.
It was found out early \cite{Anderson1952} that the groundstate 
of this model bears resemblance with the classical \textit{N\'{e}el
  state} \index{N\'{e}el antiferromagnet}
\begin{equation}
  \label{eq:neelclass}
  \ket{\text{N\'{e}el class.}} = \ket{\uparrow \downarrow \uparrow \downarrow
  \cdots}
\end{equation}
where the spin-ups live on the $A$ sublattice and the spin-downs live
on the $B$ sublattice. The total spin $S_{\text{tot}}$
is not a good quantum number for this state. From elementary spin
algebra we know that 
it is rather a superposition of several states with different total
spin quantum numbers. For example the 2-site state 
\begin{equation}
  \label{eq:singletplustriplet}
  \ket{\uparrow \downarrow } = \frac{\ket{\uparrow \downarrow} - 
    \ket{\downarrow\uparrow}}{2} + \frac{\ket{\uparrow \downarrow} + \ket{\downarrow\uparrow}}{2} 
     = \ket{S_{\text{tot}}=0, m=0} + \ket{S_{\text{tot}}=1, m=0}
\end{equation}
is the superposition of a singlet ($S_{tot}=0$) and a triplet ($S_{tot}=1$).
Therefore if such a state were to be a
groundstate of Eq.~\eqref{eq:heisenberg} several states with different
total spin would have to be degenerate. It turns out that on finite bipartite
lattices this is not the case: The total groundstate of the
Heisenberg model on bipartite lattices can be proven to be a singlet state with 
$S_{\text{tot}}=0$.
This result is known as \textit{Marshall's Theorem} 
\cite{Marshall1955,Lieb1962,Auerbach1994}.
So how can the N\'{e}el state resemble the singlet groundstate?
To understand this we drastically simplify the Heisenberg model
and investigate a toy model whose spectrum can be fully understood
analytically.

\subsection{Toy model: the Lieb-Mattis model} \index{Lieb-Mattis model}
By introducing the Fourier transformed spin operators
\begin{equation}
  \label{eq:spinoperatorfourier}
  \mathbf{S}_\mathbf{k} = \frac{1}{\sqrt{N}}\sum\limits_{j=0}^N
e^{i\mathbf{k} \cdot \mathbf{x}_j}\mathbf{S}_j
\end{equation}
we can rewrite the original Heisenberg Hamiltonian in terms of these
operators as
\begin{equation}
  \label{eq:heisenbergfourier}
  H = J\sum\limits_{\mathbf{k} \in \text{BZ}} \omega_\mathbf{k}
\mathbf{S_k}\cdot\mathbf{S}_{-\mathbf{k}}
\end{equation}
where $\omega_\mathbf{k} = \cos(k_x) + \cos(k_y)$ and the sum over $\mathbf{k}$ 
runs over the momenta within the first Brillouin zone~(BZ). Let $\mathbf{k}_0 =
(\pi,\pi)$ be the ordering wavevector which is the dual to the translations 
that leave the square N\'{e}el state invariant. We now want to look at the 
truncated Hamiltonian
\begin{equation}
  \label{eq:liebmattis}
  H_{\text{LM}} = 2J\left(\mathbf{S}_{(0,0)}^2 -
\mathbf{S}_{\mathbf{k}_0}\cdot\mathbf{S}_{-\mathbf{k}_0}\right)
\end{equation}
where we omit all Fourier components in
Eq.~\eqref{eq:heisenbergfourier} except $\mathbf{k}=(0,0)$ and
$\mathbf{k}_0 = (\pi,\pi)$.
This model is called the Lieb-Mattis model \cite{Lieb1962} and has a 
simple analytical solution. To see this we notice that
Eq.~\eqref{eq:liebmattis} can be written as  
\begin{equation}
  \label{eq:liebmattis2}
  H_{LM} = \frac{4J}{N}\sum\limits_{i\in A, j\in B} \mathbf{S}_i \cdot
\mathbf{S}_j
\end{equation}
in real space, where $A$ and $B$ denote the two bipartite sublattices of the
square lattice and each spin is only coupled with spins
in the other sublattice. The interaction strength is equal
regardless of the distance between the two spins. Thus this
model is not likely to be experimentally relevant. Yet it will serve
as an illustrative example how breaking the spin-rotational symmetry
manifests itself in the spectrum of a finite size system. 
We can rewrite Eq. \eqref{eq:liebmattis2} as
\begin{align}
  \label{eq:liebmattis3}
  H_{LM} &=  \frac{4J}{N}\left(\sum\limits_{i,j \in A\cup B} \mathbf{S}_i \cdot
\mathbf{S}_j
    - \sum\limits_{i,j \in A} \mathbf{S}_i \cdot \mathbf{S}_j -
    \sum\limits_{i,j \in B} \mathbf{S}_i \cdot \mathbf{S}_j \right) \\
  &=  \frac{4J}{N}( \mathbf{S}_{\text{tot}}^2 - \mathbf{S}_A^2 - \mathbf{S}_B^2)
\end{align}
This shows that the Lieb-Mattis model can be considered as the 
coupling of two large spins $S_A$ and $S_B$ to a total spin $S_{\text{tot}}$.

We find that the operators $\mathbf{S}_{\text{tot}}^2$, $S_{\text{tot}}^z$,
$\mathbf{S}_A^2$ and $\mathbf{S}_B^2$
commute with this Hamiltonian and therefore the sublattice spins
$S_A$ and $S_B$ as well as the total spin $S_{\text{tot}}$ and its z-component
$m_{\text{tot}}$ are good quantum numbers for this model.   
For a lattice with $N$ sites ($N$ even) the sublattice spins can be chosen in
the range $S_{A,B}\in \{0,1,\dots,N/4\}$ and by coupling them 
\begin{align}
 S_{\text{tot}} &\in \{|S_A-S_B|,|S_A-S_B|+1,\dots,S_A+S_B\} \\
 m_{\text{tot}} &\in \{-S_{\text{tot}}, -S_{\text{tot}}+1, \dots,
S_{\text{tot}}\}
\end{align}
can be chosen\footnote{This set of states spans the full Hilbertspace of the
model.}. A state $\ket{S_{\text{tot}}, m, S_A, S_B}$ is thus an eigenstate of
the systems with energy
\begin{equation}
 E(S_{\text{tot}}, m, S_A, S_B) = \frac{4J}{N}
 \left[S_{\text{tot}}(S_{\text{tot}}+1) - S_A(S_A+1) - S_B(S_B+1) \right]
 \label{eq:energy_lm}
\end{equation}
independent of $m$, so each state is at least ($2 S_{\text{tot}}+1$)-fold 
degenerate. 

\paragraph{Tower of states}
We first want to consider only the lowest energy states for each
$S_{\text{tot}}$ sector. These states build the famous \textit{tower of states}
and collapse in the thermodynamic limit to a highly degenerate groundstate 
manifold, as we will see in the following.

For a given total spin $S_{\text{tot}}$ the lowest energy states are built by
maximizing the last two terms in Eq.~\eqref{eq:energy_lm} with $S_A=S_B=N/4$ and 
\begin{equation}
 E_0(S_{\text{tot}}) = E(S_{\text{tot}},m, N/4,N/4) = \frac{4J}{N}
 S_{\text{tot}}(S_{\text{tot}}+1) - J (\frac{N}{4}+1)
 \label{eq:energy_lm1}
\end{equation}
The groundstate of a finite system will thus be the singlet state with
$S_{\text{tot}}=0$
\footnote{The groundstate of the Heisenberg model
Eq.~\eqref{eq:heisenberg} on a bipartite sublattice with equal sized sublattices
is also proven to be a singlet state $S_{tot}=0$ by Marshall's Theorem
\cite{Auerbach1994, Marshall1955, Lieb1962}.}
. On a finite system the
groundstate is, therefore, totally symmetric under global spin rotations and
does not break the $SU(2)$-symmetry. In the thermodynamic limit $N\rightarrow
\infty$, however, the energy of all these states scales to zero and all of them
constitute to the groundstate manifold. 

The classical N\'eel state with fully polarized spins on each
sublattice can be built out of these states by a linear combination of all the
$S_{\text{tot}}$ levels with $m_{tot}=0$ \cite{Lhuillier2005}. All other N\'eel
states pointing in a different direction in spin-space can be equivalently
built out of this groundstate manifold by considering linear combinations with
other $m_{tot}$ quantum numbers. In the thermodynamic limit, any infinitesimal
small field will force the N\'eel state to choose a direction and the
groundstate spontaneously breaks $SU(2)$-symmetry.

The states which constitute the groundstate manifold in the thermodynamic limit
can be readily identified on finite-size systems as well, where their energy
and spin quantum number are given by Eq.~\eqref{eq:energy_lm1}. These states
are called the \textit{tower of states} (TOS) or also \textit{Anderson tower},
\textit{thin spectrum} and \textit{quasi-degenerate joint states}
\cite{Anderson1952, Kaplan1990,Hasenfratz1993,Azaria1993}.

\paragraph{Excitations}
The lowest excitations above the tower of states can be built by lowering the
spin of one sublattice $S_A$ or $S_B$ by one, see Eq.~\eqref{eq:energy_lm}. Let
us set $S_A=N/4$ and $S_B=N/4-1$ which implies that $S_{\text{tot}} \in
\{1,2,\dots,N/2-1\}$. We can directly compute the energy $E_1(S_{\text{tot}})$
of these excited states for each allowed $S_{\text{tot}}$. The energy gap to
the tower of states
\begin{equation}
 E_{\text{exc}}(S_{\text{tot}}) = E_1(S_{\text{tot}}) - E_0(S_{\text{tot}}) = J
\end{equation}
is constant\footnote{This is an artifact of the infinite-range interaction in
the Lieb-Mattis model. In the original Heisenberg model these modes become
gapless magnon excitations.}.
Hence, the lowest excitations of the Lieb-Mattis model
are static spinflips. The next lowest excitations are spinflips on
both sublattices, $S_A=S_B=N/4-1$ with excitation
energy $E_{\text{exc}_2} = 2J$ and $S_{\text{tot}} \in \{0,1,\dots,N/2-2\}$. We
observe that only the energy gap of the TOS levels vanishes
in the thermodynamic limit, so the TOS indeed solely contributes to the
groundstate manifold.

\paragraph{Quantum Fluctuations}
When we introduced the Lieb-Mattis model Eq.~\eqref{eq:liebmattis} from the
Heisenberg model Eq.~\eqref{eq:heisenbergfourier} we neglected all Fourier
components except of $\mathbf{k}=(0,0)$ and $\mathbf{k} = \mathbf{k}_0$. This
was a quite crude approximation and it is not guaranteed that all results for
the Lieb-Mattis model will survive for the short-range Heisenberg model. To get
some first results regarding this question, we can introduce small quantum
fluctuations on top of the N\'eel groundstate of the Lieb-Mattis model and
perform a perturbative spin-wave analysis in first order\footnote{A more
detailed discussion can be found in \cite{Lhuillier2005}.}. This approach does
not affect the scaling of the tower of states levels, but it has an important
effect on the excitations. They are not static particles anymore, but become
spinwaves (magnons) with a dispersion, which is linear around the ordering-wave
vector $\mathbf{k}=\mathbf{k}_0$ and $\mathbf{k}= (0,0)$. On a finite-size
lattice the momentum space is discrete with a distance proportional to $1/L$
between momentum space points, where $L$ is the linear size of the system. The
energy of the lowest excitation above the TOS, the single magnon gap, therefore
scales as $E_{\text{exc}} \propto J/L$ to zero\footnote{In the thermodynamic
    limit the single magnon mode is gapless and has linear dispersion around
    $\mathbf{k}=\mathbf{k}_0$ and $\mathbf{k}= (0,0)$. It corresponds to the
    well-known Goldstone mode which is generated when a continuous symmetry is
spontaneously broken.}.
As the scaling is, however, slower for $d>1$-dimensional systems than the TOS
scaling, these levels do not influence the groundstate manifold in the
thermodynamic limit. Furthermore, the excitation of two magnons results in a
two-particle continuum above the magnon mode.

\begin{figure}[ht]
 \centering
   \includegraphics[width=.4\textwidth]{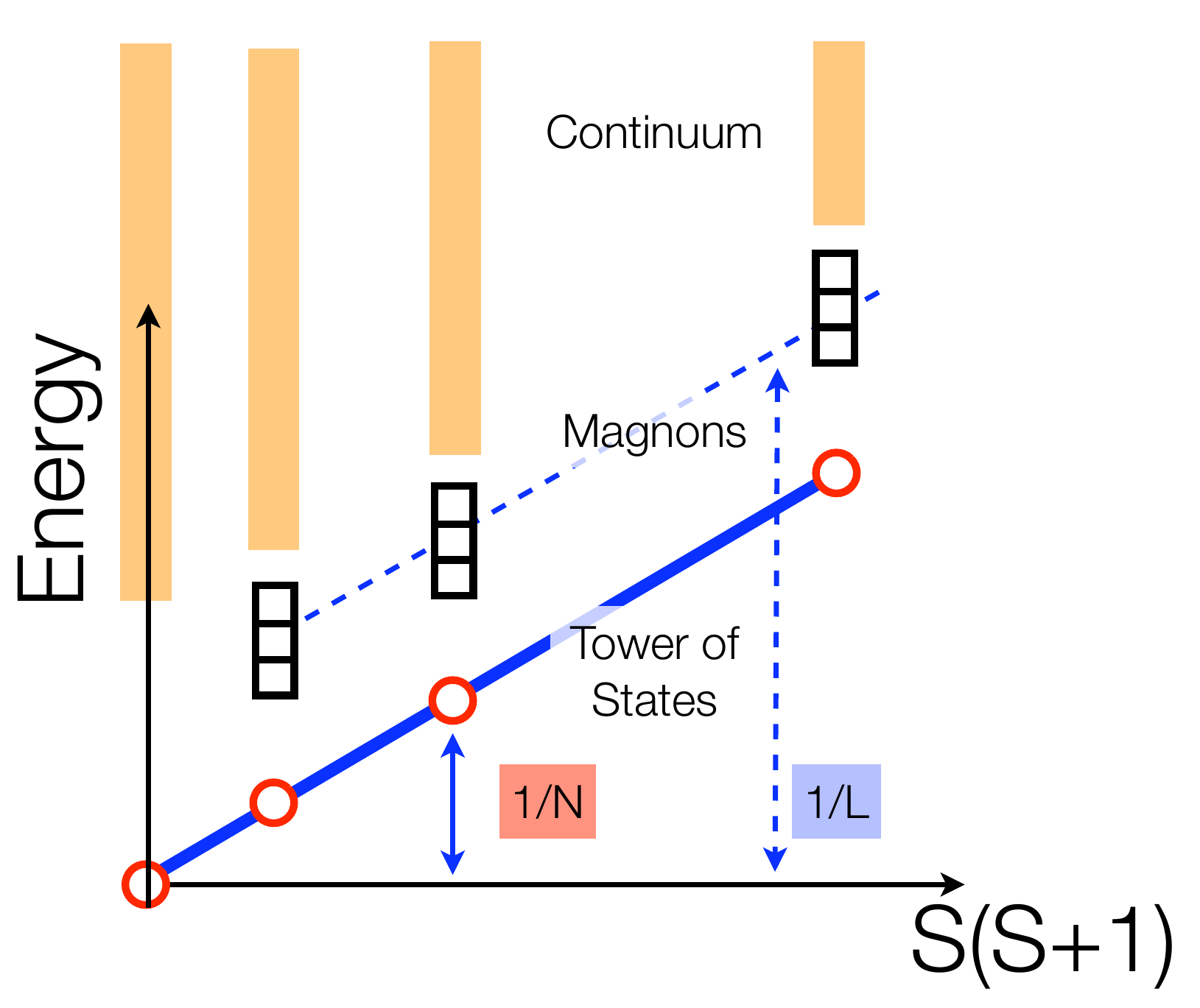}\hspace{10mm}$\mbox{}$
  \includegraphics[width=.4\textwidth]{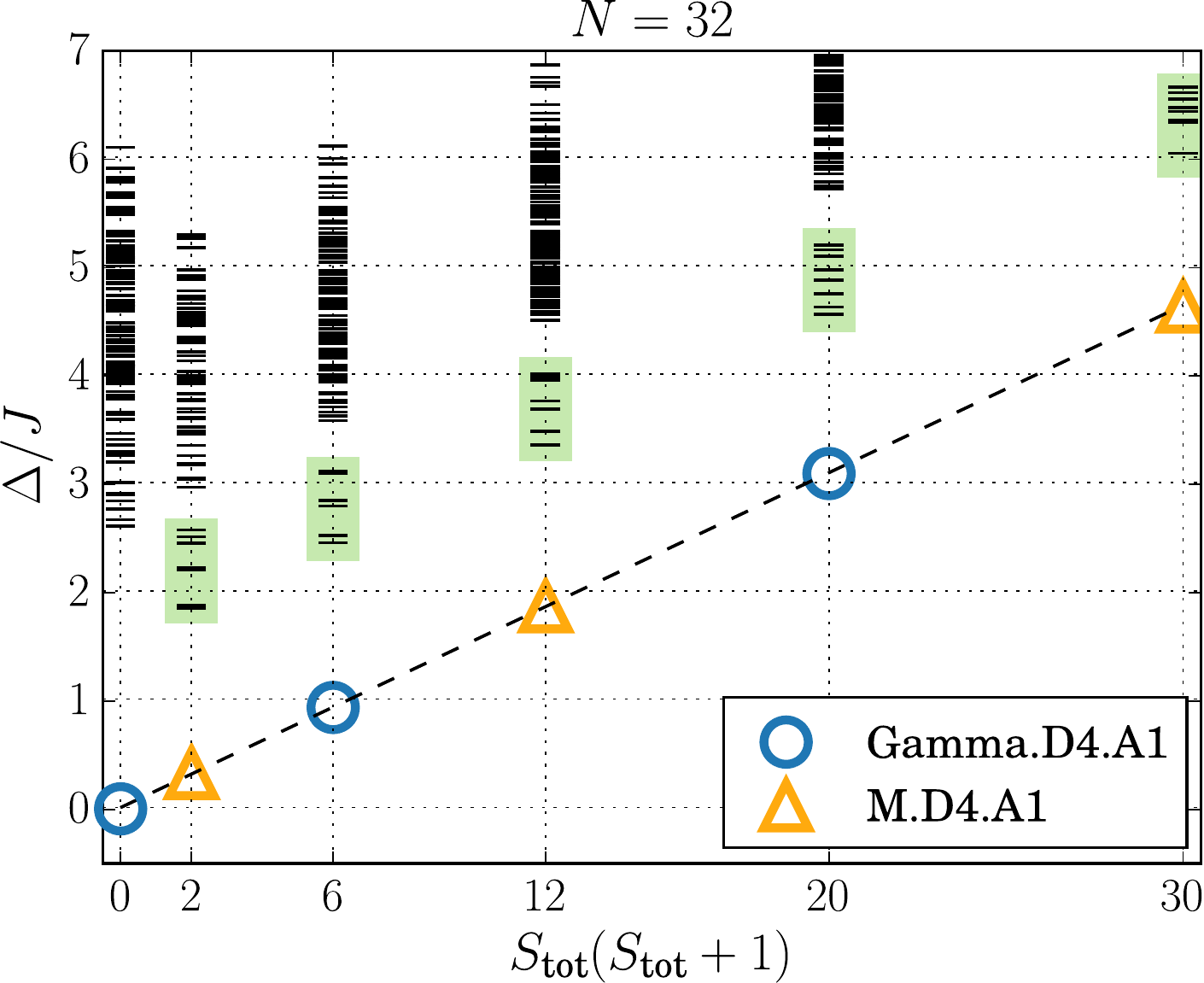}
  \caption{(Left): Schematic finite-size energy spectrum of an antiferromagnet
   breaking SU($2$) spin-rotational symmetry. The TOS levels are the
lowest energy levels for each total spin $S$ and scale with
$1/N$ to the groundstate energy. The low energy magnon
excitations are seperated from the TOS and a continuum of higher
energy states and scale with $1/L$. (Right): Energy spectrum for the
Heisenberg model on a square lattice from ED. The TOS levels are connected by a dashed
line. The single magnon dispersion (green boxes) with $S_{\text{tot}} \in
\{1,2,\dots\}$ are well separated from the TOS and the higher multi-particle
continuum. The different symbols represent quantum numbers related to
space-group symmetries and agree with the expectations for a N\'eel state
(See section~\ref{sec:sym_analysis}).}
 \label{fig:TOS_Neel}
\end{figure}

The properties of the TOS and its excitations are summarized in
Fig.~\ref{fig:TOS_Neel}. The left figure shows the general properties of the
finite-size energy spectrum which can be expected when a continuous symmetry
group is spontaneously broken in the thermodynamic limit. The right figure
depicts the energy spectrum for the Heisenberg model on a square lattice with
$N=32$ sites, obtained with Exact Diagonalization. One can clearly identify the
TOS, the magnon dispersion and the many-particle continuum. The existence of a
N\'eel TOS was not only confirmed numerically for the Heisenberg model on the
square lattice, but also with analytical techniques beyond the simplification
to the Lieb-Mattis model \cite{Anderson1952,Hasenfratz1993,Azaria1993}. The
different symbols in Fig.~\ref{fig:TOS_Neel} represent different quantum numbers
related to the space-group symmetries on the lattice. In the next section we
will see that the structure of these quantum numbers depends on the exact shape
of the symmetry-broken state and we will learn how to compute them. 

\section{Symmetry analysis} \label{sec:sym_analysis}

In the analysis of excitation spectra from Exact Diagonalization on
finite size simulation clusters the TOS analysis is a powerful tool
to detect spontaneous symmetry breaking. As we have seen in the previous
chapter explicitly for the Heisenberg antiferromagnet,
symmetry breaking implies degenerate groundstates in the
thermodynamic limit. On finite size simulation clusters this degeneracy
is in general not an exact degeneracy.  We rather expect a certain
scaling of the energy differences in the thermodynamic limit.
We distinguish two cases:

\begin{itemize}
\item \textbf{Discrete symmetry breaking:} In this case we have
  a degeneracy of finitely many states in the thermodynamic limit. 
  The groundstate splitting $\Delta$ on finite size clusters 
  scales as $\Delta \sim \exp(-N/\xi)$, where $N$ is the number of
  sites in the system.
\item \textbf{Continuous symmetry breaking:} The groundstate
  in the thermodynamic limit is infinitely degenerate. The states
  belonging to this degenerate manifold collapse as \mbox{$\Delta \sim
  1/N$} on finite size clusters as we have seen in section~\ref{sec:TOS}.
  It is important to understand that these states
  are not the Goldstone modes of continuous symmetry breaking.
  Both the degenerate groundstate and the Goldstone modes 
  appear as low energy levels on finite size clusters but have
  different scaling behaviours.
\end{itemize}

The scaling of these low energy states can now be investigated on finite size
clusters. More importantly also their quantum numbers such as momentum,
pointgroup representation or total spin can be predicted
\cite{Lhuillier2005,Misguich2007,Rousochatzakis2008}. 
The detection of correct scaling 
behaviour together with correctly predicted quantum numbers 
yields very strong evidence that the system spontaneously breaks
symmetry in the way that has been anticipated. This is the TOS 
method. In the following we will discuss how to predict the 
quantum numbers for discrete as well as continuous symmetry breaking.
The main mathematical tool we use is the character-formula
from basic group representation theory. 

Lattice Hamiltonians like a Heisenberg model often have
a discrete symmetry group arising from translational invariance,
pointgroup invariance or some discrete local symmetry, like a 
spinflip symmetry. In this chapter we will first discuss the
representation theory and the characters of the representations
of space groups on finite lattices. We will then see how 
this helps us to predict the representations of the 
degenerate ground states in discrete as well as continuous symmetry breaking.

\subsection{Representation theory for space groups} 
\index{representation theory}
For finite discrete groups such as the space group
of a finite lattice the full set of irreducible representations 
(irreps) can be worked out. 
Let us first discuss some basic groups.
Let's consider a $n\times n$ square lattice with periodic 
boundary conditions and a translationally invariant Hamiltonian
like the Heisenberg model on it. In the following we will 
set the lattice spacing to $a=1$. The discrete symmetry
group we consider is $\mathcal{T} = \mathbb{Z}_n \times \mathbb{Z}_n$
corresponding to the group of translations on this lattice. 
This is an Abelian group of order $n^2$. Its representations
can be labeled by the momentum vectors $\mathbf{k} = (\frac{2\pi
  i}{n}, \frac{2\pi j}{n})$, $i,j \in \{ 0, \cdots, n-1\} $
which just correspond to the reciprocal Bloch vectors defined on 
this lattice. Put differently, the vectors $\mathbf{k}$ are the reciprocal lattice points
of the lattice spanned by the simulation torus of our $n\times n$ square lattice.
The character $\chi_\mathbf{k}$ of the $\mathbf{k}$-representation is given by 
\begin{equation}
  \label{eq:transchars}
  \chi_\mathbf{k}(\mathbf{t}) = \text{e}^{i\mathbf{k}\cdot \mathbf{t}}
\end{equation}
where $\mathbf{t} \in \mathcal{T}$ is the vector of translation. This is just 
the usual Bloch factor for translationally invariant systems. 

Let us now consider a (symmorphic) space group of the form
$\mathcal{D} = \mathcal{T}\times \text{PG}$ as 
the discrete symmetry group of the lattice 
where $\text{PG}$ is the pointgroup of the lattice. For a model on
a $n\times n $ square lattice this could for example be the dihedral
group of order 8, D$_4$, consisting of fourfold rotations
together with reflections. The representation theory and 
the character tables of these point groups are well-established.
An example for such character tables can be found in
Tabs.~\ref{tab:char_table_c4} and \ref{tab:d6chartable} for the 
cyclic group $C_4$ and the dihedral group $D_6$~\footnote{We follow the
labeling scheme for point group representations according to
Mulliken~\cite{Mulliken1955}.}.
Since $\mathcal{D}$ is now a product of the translation and
the point group we could
think that the irreducible representations of $\mathcal{D}$
are simply given by the product representations $(\mathbf{k} \otimes \rho)$
where $\mathbf{k}$ labels a momentum representation and $\rho$ an irrep 
of $\text{PG}$. But here is a small yet important caveat. We have to be careful
since $\mathcal{D}$ is only a semidirect product of groups as 
translations and pointgroup symmetries do not necessarily commute.
This alters the representation theory for this product of groups
and the irreps of $\mathcal{D}$ are not just simply the products
of irreps of $\mathcal{T}$ and $\text{PG}$. Instead the full set
of irreps for this group is given by $(\mathbf{k} \otimes \rho_\mathbf{k})$ where
$\rho_\mathbf{k}$ is an irrep of the so called \textit{little group} 
$L_\mathbf{k}$ of $\mathbf{k}$ defined as
\index{representation theory:little group}
\begin{equation}
  \label{eq:deflittlegroup}
  L_\mathbf{k} = \left\{ g \in \text{PG}; g(\mathbf{k}) = \mathbf{k} \right\}
\end{equation}
which is just the stabilizer of $\mathbf{k}$ in $\text{PG}$. For example, all
pointgroup elements leave $\mathbf{k}=(0,0)$ invariant, thus the little group
of $\mathbf{k}=(0,0)$ is the full pointgroup PG. In general, this does not hold 
for other momenta and only a subgroup of  $\text{PG}$ will be 
the little group of $\mathbf{k}$. In Fig.~\ref{fig:3sublatt_triangular}
we show the $\mathbf{k}$-points of a $6\times6 $ triangular 
lattice together with its little groups as an example. The $K$ point
in the Brillouin zone has a D$_3$ little group, the $M$ point a D$_2$ 
little group. Having discussed the represenation theory for 
(symmorphic) space groups we state that the characters of these
representations are simply given by
\begin{equation}
  \label{eq:spacechars}
  \chi_{(\mathbf{k},\rho_\mathbf{k})}(\mathbf{t},p) = \text{e}^{i \mathbf{k}\cdot \mathbf{t}} \chi_{\rho_\mathbf{k}}(p)
\end{equation}
where $\mathbf{t} \in \mathcal{T}$, $p \in \text{PG}$ and $\chi_{\rho_\mathbf{k}}$ denotes
the character of the representation $\rho_\mathbf{k}$ of the little group
$L_\mathbf{k}$. 

\subsection[Predicting irreducible representations in spontaneous 
  symmetry breaking]{Predicting irreducible representations in spontaneous 
  symmetry breaking \subsectionmark{Predicting irreducible representations}
}
\subsectionmark{Predicting irreducible representations}
\index{tower of states!quantum numbers}
\index{representation theory:irreducible representations}
Spontaneous symmetry breaking at $T=0$ occurs when 
the groundstate $\ket{\psi_{\textnormal{GS}}}$ of $H$ in the
thermodynamic limit is not invariant under the full symmetry group
$\mathcal{G}$ of $H$. We will call a specific groundstate
$\ket{\psi_{\textnormal{GS}}}$ a \textit{prototypical state} and  
the \textit{groundstate manifold} is defined by 
\begin{equation}
  \label{eq:defdegenerategs}
  V_{\textnormal{GS}} = \spn\left\{ \ket{\psi_{\textnormal{GS}}^i }\right\}
\end{equation}
where $\ket{\psi_{\textnormal{GS}}^i }$ is the set of degenerate
groundstates in the thermodynamic limit.
This groundstate manifold space can be finite or infinite dimensional
depending on the situation. For breaking a discrete finite symmetry, 
such as in the example given in section \ref{sec:sym_analysis_VB}, 
this groundstate manifold will be finite dimensional, for breaking continuous SO($3$) 
spin rotational symmetry\footnote{The actual symmetry group of
  Heisenberg antiferromagnets is usually SU($2$). For simplicity
  we only consider the subgroup SO($3$) in these notes which yields
  the same predictions for the case of sublattices with even number of 
  sites (corresponding to integer total sublattice spin).} 
as in section~\ref{sec:cont_symm_breaking}
it is infinite dimensional in the thermodynamic
limit. For every symmetry $g \in \mathcal{G}$ 
we denote by $O_g$ the symmetry operator acting on the Hilbert space. 
The groundstate manifold becomes degenerate in the thermodynamic 
limit and we want to calculate the quantum numbers of the groundstates
in this manifold. Another way of saying this is that we want to compute
the irreducible representations of $\mathcal{G}$ to which the groundstates belong
to. 
For this we look at the action $\Gamma$ of the symmetry group $\mathcal{G}$ on 
$V_{\textnormal{GS}}$ defined by 
\begin{align}
  \label{eq:representationongroundstates}
  \Gamma: &\mathcal{G} \rightarrow \aut(V_{\textnormal{GS}})\\
  & g \mapsto \left(  \braket{\psi_{\textnormal{GS}}^i| O_g |\psi_{\textnormal{GS}}^j}\right)_{i,j}
\end{align}
This is a representation of $\mathcal{G}$ on $V_{\textnormal{GS}}$, so
every group element $g \in \mathcal{G}$ is mapped to an invertible
matrix on $V_{\textnormal{GS}}$. In general this representation is
reducible and can be decomposed into a direct sum of irreducible representations
\begin{equation}
  \label{eq:ssbirrepdecomp}
  \Gamma = \bigoplus_\rho n_\rho \rho
\end{equation}
These irreducible representations $\rho$ are the quantum
numbers of the eigenstates in the groundstate manifold and $n_\rho$ 
are its respective multiplicities (or degeneracies). 
Therefore these irreps constitute the TOS for spontaneous symmetry
breaking \cite{Lhuillier2005}. To compute the
multiplicities we can use a central result from representation theory,
the \textit{character formula}
\begin{equation}
  \label{eq:characterformula}
  n_\rho = \frac{1}{|\mathcal{G}|}\sum\limits_{g\in \mathcal{G}}\overline{\chi_\rho(g)}\tr(\Gamma(g))
\end{equation}
where $\chi_\rho(g)$ is the character of the representation $\rho$ and
$\tr(\Gamma(g))$ denotes the trace over the representation matrix 
$\Gamma(g)$ as defined in Eq.~\eqref{eq:representationongroundstates}.
Often we have the case that
\begin{equation}
  \label{eq:vanishingoverlapthermolim}
   \braket{\psi_{\textnormal{GS}} | O_g | \psi_{\textnormal{GS}}'} = 
   \begin{cases}
     1 \text{ if } O_g\ket{\psi_{\textnormal{GS}}'} = \ket{\psi_{\textnormal{GS}}} \\
     0 \text{ otherwise}
   \end{cases} 
\end{equation}
With this we can simplify Eq.~\eqref{eq:characterformula} to 
what we call the \textit{character-stabilizer formula}
\begin{equation}
  \label{eq:stabilizerformula}
  n_{\rho} = \frac{1}{|\text{Stab}(\ket{\psi_{\textnormal{GS}}})|}\sum
  \limits_{g \in \text{Stab}(\ket{\psi_{\textnormal{GS}}})}\chi_\rho(g)
\end{equation}
where 
\begin{equation}
  \label{eq:stabilizerdefinition}
  \text{Stab}(\ket{\psi_{\textnormal{GS}}}) \equiv \{ g \in \mathcal{G} :
  \; O_g
  \ket{\psi_{\textnormal{GS}}} = \ket{\psi_{\textnormal{GS}}}\}
\end{equation}
is the stabilizer of a prototypical state
$\ket{\psi_{\textnormal{GS}}}$
~\footnote{In some cases, the orbit of the prototypical state 
{$G.\ket{\psi_{GS}} = \{g \in G: \; O_g \ket{\psi_{GS}}\}$} does
not span the full set of degenerate groundstates
{$\ket{\psi^i_{GS}}$}.
In this case, we have to find a set of prototypical states with
different orbits, such that the union of these orbits spans the
full groundstate manifold. Then, Eq.~\unexpanded{\eqref{eq:stabilizerformula}}
has to be applied to each prototypical state, individually, and
the final multiplicity is the sum of the individual results.
}. 
We see that for applying the character-stabilizer formula in 
Eq.~\eqref{eq:stabilizerformula} only two ingredients are needed:
\begin{itemize}
\item the stabilizer
  $\text{Stab}(\ket{\psi_{\textnormal{GS}}})$ of a prototypical state
  $\ket{\psi_{\textnormal{GS}}}$ in the groundstate manifold
\item the characters of the irreducible representations of the
  symmetry group $\mathcal{G}$
\end{itemize}
We want to remark that in the case
of $\mathcal{G} =  \mathcal{D} \times \mathcal{C}$ 
where $\mathcal{D}$ is a discrete symmetry group such as the spacegroup of a
lattice and $\mathcal{C}$ is a continuous symmetry group such as
SO($3$) rotations for Heisenberg spins the 
Eqs.~\eqref{eq:characterformula} and \eqref{eq:stabilizerformula}
include integrals over Lie groups additionally to the sum over 
the elements of the discrete symmetry group $\mathcal{D}$.
Furthermore also the characters for Lie groups like SO($3$) 
are well-known. For an element $R \in $ SO($3$) the irreducible
representations are labeled by the spin $S$ and its characters
are given by 
\begin{equation}
       \chi_S(R) = \frac{\sin\left[(S+\frac{1}{2})\phi\right]}{\sin\frac{\phi}{2}}
\end{equation}
where $\phi \in [0, 2\pi]$ is the angle of rotation of the spin
rotation $R$. We work out several examples for this case in 
section~\ref{sec:cont_symm_breaking}
and compare the results to actual numerical data from Exact Diagonalization.

\section{Examples}\label{sec:examples}
\subsection{Discrete symmetry breaking}

In this section we want to apply the formalism of 
section~\ref{sec:sym_analysis} to
systems, where only a discrete symmetry group is spontaneously broken but not a
continuous one. In this case, the ground-state of the system in the thermodynamic
limit is described by a superposition of a finite number of degenerate
eigenstates with different quantum numbers. On finite-size systems, however,
the symmetry cannot be broken spontaneously and a unique groundstate will be
found. The other states constituting to the degenerate eigenspace in the
thermodynamic limit exhibit a finite-size energy gap which is exponentially
small in the system size $N$, $\Delta \propto \mbox{e}^{-N/\xi}$. The quantum
numbers of these quasi-degenerate set of eigenstates are defined by the
symmetry-broken state in the thermodynamic limit.

\subsubsection{Introduction to valence-bond solids}\index{valence-bond solid}

In section~\ref{sec:TOS} we have seen that the classically ordered
N\'eel state is a candidate to describe the groundstate of the
antiferromagnetic Heisenberg model Eq.~\eqref{eq:heisenberg}
with $J>0$ in the thermodynamic limit on a bipartite lattice. The energy
expectation value of this state on a single bond is $e_{\text{N\'eel}} = -J/4$. 

The state which minimizes the energy of a single bond is, however, a singlet
state $\ket{S=0}$ formed by the two spins on the bond with energy 
$e_{\text{VB}} = -3J/4$, called a valence bond (VB) or dimer. A valence bond 
covering of an $N$-site lattice can then be described by a tensor product of 
$N/2$ VBs, where each site belongs to exactly one VB\footnote{The set of all 
possible valence bond coverings with arbitrary length spans the full 
$S_{\text{tot}}=0$ sector of the models Hilbert space and is overcomplete 
{\cite{Liang1988,Lhuillier2011}}.}.
Another possible candidate for the thermodynamic groundstate of 
Eq.~\eqref{eq:heisenberg} is then a superposition of all possible VB coverings 
with only nearest neighbour VBs. Such states do not break the $SU(2)$ 
spin-rotational symmetry as $S_{\text{tot}}=0$ and are in general not 
eigenstates of the Hamiltonian: Acting with the operator $\mathbf{S}_i \cdot
\mathbf{S}_j$ between sites $i$ and $j$ belonging to two different VBs changes 
the VB configuration. 

This manifold of VB coverings is highly degenerate. As the VB coverings
are in general not eigenstates of the Hamiltonian, they encounter quantum
fluctuations. The energy corrections due to these fluctuations are typically
not equivalent for different coverings, although the bare energies are 
identical.
The VB coverings with the largest energy gain are selected by the fluctuations
as the true groundstate configurations. If this \textit{order-by-disorder}
mechanism \cite{Shender1982,Henley1989} selects regular patterns \index{order by disorder}
of VB coverings, the discrete lattice symmetries can be spontaneously broken in the
thermodynamic limit, and a \textit{valence bond solid} (VBS)\index{valence bond 
solid} can be
formed. Fig.~\ref{fig:cvbs} and Fig.~\ref{fig:svbs} show two different VBS
states on the square lattice. VBSs show no long-range spin order, but long-range
dimer-correlations $\langle(\mathbf{S}_a \cdot \mathbf{S}_{a'}) (\mathbf{S}_b
\cdot \mathbf{S}_{b'})\rangle$ where $a,a'$ and $b,b'$ label sites on
individual dimers. In section~\ref{sec:sym_analysis_VB} we will see how 
different VBS states can be identified and distinguished by the quantum numbers 
of the quasi-degenerate groundstate manifold on finite-size systems.

The groundstate of the Heisenberg model Eq.~\eqref{eq:heisenberg} on the square
lattice is not a VBS but a N\'eel state, which has a lower variational energy
already on the classical level. Nevertheless, several models featuring VBS
groundstates are known in 1- and 2-D
\cite{Fouet2001,Lauchli2002,Lauchli2005,Mambrini2006,Gelle2008}. Interestingly,
in \cite{Sandvik2007} a model was proposed, which shows a direct continuous
quantum phase transition between a N\'eel state and a VBS. This transition
exhibits very exotic, non-classical behaviour and is called \textit{deconfined
quantum critical point} \cite{Senthil2004}.

\subsubsection{Identification of VBSs from finite-size spectra} 
\label{sec:sym_analysis_VB}

\begin{figure}[h]
 \centering
 \includegraphics[width=\textwidth]{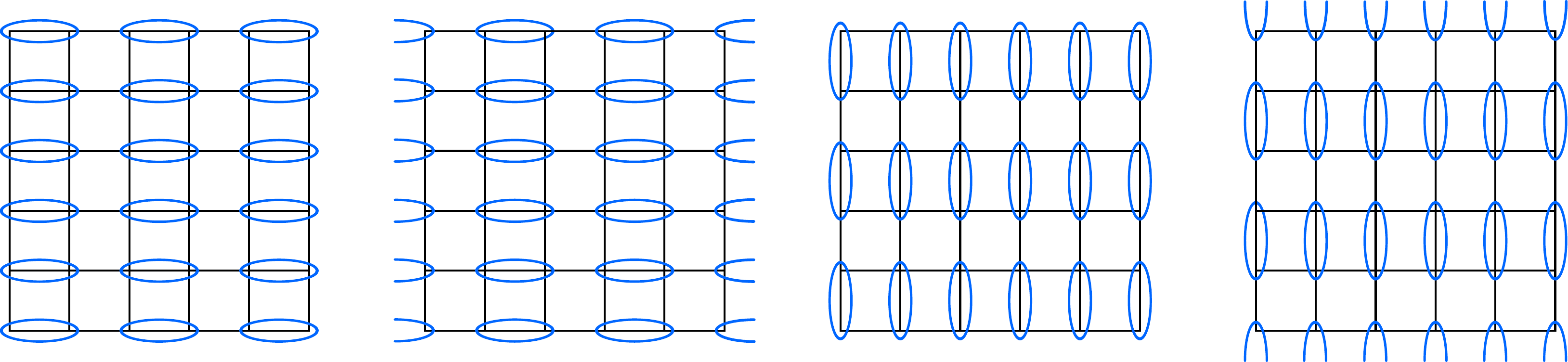}
 \caption{The four columnar VBS coverings of a square lattice. Valence bonds 
(spin singlets) are indicated by blue ellipses.}
 \label{fig:cvbs}
\end{figure}

\paragraph{Columnar valence-bond solid}
\index{valence-bond solid!columnar}
A columnar VBS (cVBS) on a square lattice is shown in Fig.~\ref{fig:cvbs}. Four
equivalent states can be found, indicating that there will be a four-fold
quasi-degenerate groundstate manifold. A cVBS breaks the translational
and point-group symmetries of an isotropic SU(2)-invariant Hamiltonian on the
lattice spontaneously but not the continuous spin symmetry group since it
is a singlet and thus invariant under spin rotations.

In the following we use Eq.~\eqref{eq:stabilizerformula} to compute the
symmetry sectors of the groundstate manifold. The discrete symmetry group
we consider is
\begin{equation}
 \mathcal{G} = \mathcal{D} = \mathcal{T} \times \text{PG}
\end{equation}
where $\mathcal{T} = \mathbb{Z}_2 \times \mathbb{Z}_2 = \{1, t_x, t_y,
t_x t_y\}$ are the non-trivial lattice translations with translation vectors
\begin{equation}
 \mathbf{t}_1 = (0,0), \quad \mathbf{t}_x = (1,0), \quad 
\mathbf{t}_y= (0,1), \quad \mathbf{t}_{xy} = (1,1)
\end{equation}
and $\text{PG} = \text{C}_4$ denotes the point-group of four-fold lattice
rotations\footnote{The dihedral group $\text{D}_4$ is also a symmetry group of
the model. For the sake of simplicity we decided to only consider the subgroup
$\text{C}_4$ in this section.}. To compute the groundstate symmetry sectors we
do not need to consider the full symmetry group $\mathcal{G}$ but only the
stabilizer $\text{Stab}(\ket{\Psi_{cVBS}})$, leaving one of the states in
Fig.~\ref{fig:cvbs} unchanged. Without loss of generality we choose the first
covering as prototype 
$\ket{\Psi_{cVBS}}$. The stabilizer is given by
\begin{equation}
 \text{Stab}(\ket{\Psi_{cVBS}}) = \{1\times1\} \cup \{1\times C_2\} \cup
\{t_y \times 1\} \cup \{t_y \times C_2\}
\end{equation}
where $C_2$ denotes the rotation about an angle $\pi$ around the center of a
plaquette.

The irreducible representations (irreps) of the group of lattice translations 
$\mathcal{T}$ can be labelled by the allowed momenta $\mathbf{k}$
\begin{equation}
 \mathbf{k} \in \mbox{Irreps}(\mathcal{T}) = \{(0,0), (\pi,0), (0,\pi),
(\pi,\pi)\},
\end{equation}
and the corresponding characters for an element $t \in \mathcal{T}$ are
\begin{equation}
 \chi_{\mathbf{k}}(t) = \mathrm{e}^{i \mathbf{k} \cdot \mathbf{t}}.
\end{equation}
The irreps (called A, B and E, see~\cite{Mulliken1955}) and
characters for the point-group
$\text{C}_4$ are tabulated in Tab.~\ref{tab:char_table_c4}.

\begin{table}[htb]
 \begin{center}
 \begin{tabular}{c|c|c|c|c}
  $\text{C}_4$ & $1$ & $C_4$ & $C_2$ & $(C_4)^3$ \\ \hline
  A & +1 & +1 & +1 & +1 \\
  B & +1 & -1 & +1 & -1 \\
  E$_a$ & +1 & +i & -1 & -i \\
  E$_b$ & +1 & -i & -1 & +i
 \end{tabular}
 \caption{Character table for pointgroup $\mathbf{C}_4$.}
 \label{tab:char_table_c4}
 \end{center}
\end{table}

Using the character-stabilizer formula Eq.~\eqref{eq:stabilizerformula}
we can now reduce the representation induced by the state $\ket{\Psi_{cVBS}}$
to irreducible representations to get the
quantum numbers of the quasi-degenerate groundstate manifold. Let us explicitely
consider $\mathbf{k}=(0,0)$ as an example:
\begin{align}
 n_{(0,0)A} &= n_{(0,0)B} = \frac{1}{|\text{Stab}(\ket{\Psi_{cVBS}})|} \sum_{d \in
 \text{Stab}(\ket{\Psi_{cVBS}})}
 \chi_{A}(d) \chi_{\mathbf{k}=(0,0)}(d) \\
 &= \frac{1}{4} \left[ 1\cdot  e^{i \mathbf{k} \cdot (0,0)} + 
 1\cdot  e^{i \mathbf{k} \cdot (0,0)} +
 1\cdot  e^{i \mathbf{k} \cdot (0,1)} + 
 1\cdot  e^{i \mathbf{k} \cdot (0,1)} \right] = 1 \\
 n_{(0,0)E_a} &= n_{(0,0)E_b} = \frac{1}{|\text{Stab}(\ket{\Psi_{cVBS}})|} \sum_{d \in
 \text{Stab}(\ket{\Psi_{cVBS}})}
 \chi_{B}(d) \chi_{\mathbf{k}=(0,0)}(d) \\
 &= \frac{1}{4} \left[ 1\cdot  e^{i \mathbf{k} \cdot (0,0)} + 
 (-1)\cdot  e^{i \mathbf{k} \cdot (0,0)} +
 1\cdot  e^{i \mathbf{k} \cdot (0,1)} + 
 (-1)\cdot  e^{i \mathbf{k} \cdot (0,1)} \right] = 0 
\end{align}

Eventually, the cVBS covering will be described by a four-fold quasi-degenerate
groundstate manifold with the following quantum numbers\footnote{The little
group $\rho_{\mathbf{k}}$ for the momenta $\mathbf{k}=(0,\pi)$ and
$\mathbf{k}=(\pi,0)$ is only the subgroup C$_2$ of 2-fold roations of the
symmetry group C$_4$ considered in this example. The irrep called A therefore
denotes the trivial irreducible representation of C$_2$ for these momenta. When
computing the multiplicities for these momenta one should also note, that two
prototype states have to be considered to span the full groundstate manifold
under the symmetry elements of C$_2$.}. 
\begin{equation}
 \chi(\ket{\Psi_{cVBS}}) = (0,0) \mbox{A} \oplus (0,0) \mbox{B} \oplus (\pi,0) 
\mbox{A}
\oplus (0,\pi) \mbox{A}.
\end{equation}
VBS states are a superposition of spin singlets on the lattice, therefore the
spin quantum number for all levels in the groundstate manifold must be trivial,
$S_{\text{tot}}=0$.

\begin{figure}[htb]
 \centering
 \includegraphics[width=.25\textwidth]{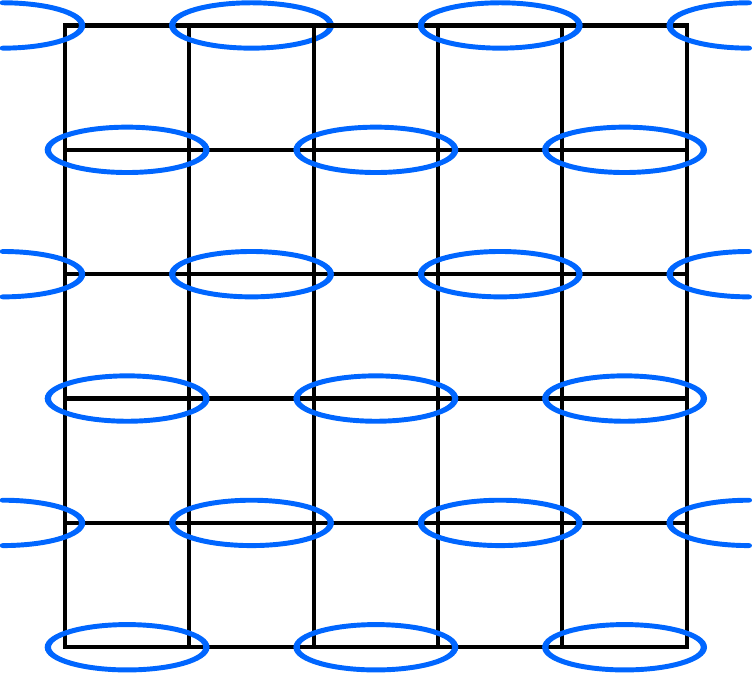}
 \caption{One of the four identical staggered VBS coverings on the square
lattice.}
 \label{fig:svbs}
\end{figure}

\paragraph{Staggered valence-bond solid}
\index{valence-bond solid!staggered}
The columnar VBS is not the only regular dimer covering of the square lattice.
Another possible regular covering is the staggered VBS (sVBS), where again four
equivalent configurations span the groundstate manifold. One of these
configurations is shown in Fig.~\ref{fig:svbs}.

Similarly, also the sVBS spontaneously breaks the translational and point-group
symmetries of an isotropic Hamiltonian, but not the spin-rotational symmetry.
Following the same steps as before we can compute the quantum
numbers of the four quasi-degenerate groundstates for the sVBS. The stabilizer
turns out to be different to the case of the cVBS and thus also the
decomposition into irreps yields a different result:
\begin{equation}
 \chi(\ket{\Psi_{sVBS}}) = (0,0) \mbox{A} \oplus (0,0) \mbox{B} \oplus
 (\pi,\pi) \mbox{E}_a \oplus  (\pi,\pi) \mbox{E}_b.
\end{equation}

Tab.~\ref{tab:irreps_VBS} shows a comparison of the irreducible representations 
in the groundstate manifold of the cVBS and sVBS states.

\begin{table}[ht]
\centering
\begin{tabular}{c|cc}
Irreps & cVBS & sVBS\\ \hline
$(0,0)$ A & 1 & 1\\
$(0,0)$ B & 1 & 1\\
$(\pi,0)$ A & 1 & 0\\
$(0,\pi)$ A & 1 & 0\\
$(\pi,\pi) \text{E}_a$  & 0 & 1\\
$(\pi,\pi) \text{E}_b$  & 0 & 1\\
\end{tabular}
\caption{Multiplicities of the irreducible representations in the four-fold
degenerate groundstate manifolds of the columnar and staggered VBS on a square
lattice.}
\label{tab:irreps_VBS}
\end{table}

By a careful analysis of the quasi-degenerate states and their quantum numbers
on finite systems it is thus possible to identify and distinguish different
VBS phases which spontaneously break the space group
symmetries in the thermodynamic limit.

\subsection{Continuous symmetry breaking}\label{sec:cont_symm_breaking}
In this section we give several examples of systems breaking continuous
SO($3$) symmetry. We discuss the introductory example of the square
lattice Heisenberg antiferromagnet, calculate the irreps in the TOS and
compare this to actual energy spectra from Exact Diagonalization on a
finite lattice in section~\ref{sec:squareantif}. 
In section~\ref{sec:magtri} we discuss three magnetic orders on the
triangular lattice and an extended Heisenberg model where all of these are
stabilized. We present results from Exact Diagonalization and compare the
representations in these spectra to the predictions from TOS analysis.
Finally, we introduce quadrupolar order and show that also this kind
of symmetry breaking can be analyzed using the TOS technique in 
section~\ref{sec:quadru}.

\subsubsection{Heisenberg antiferromagnet on the square lattice}
\label{sec:squareantif}
\index{Heisenberg model}\index{antiferromagnetism}
We now give a first example how the TOS method
can be applied to predict the structure of the tower of states 
for magnetically ordered phases. We look at the N\'{e}el
state of the antiferromagnet on the bipartite square lattice
with sublattices $A$ and $B$. 
A prototypical state in the  groundstate manifold is given by
\begin{equation}
  \label{eq:neel}
  \ket{\psi} =\ket{\uparrow\downarrow\uparrow\downarrow \cdots}
\end{equation}
where all spins point up on sublattice $A$ and down on sublattice $B$.
The symmetry group $\mathcal{G} = \mathcal{D} \times \mathcal{C}$  of 
the model we consider is a product between discrete
translational symmetry $\mathcal{D} =\mathbb{Z}_2\times \mathbb{Z}_2 =
\left\{ 1, t_x, t_y,  t_{xy}\right\}$ and spin rotational symmetry 
$\mathcal{C} = \text{SO(3)}$.
We remark that we restrict our translational symmetry group to
$\mathcal{D} = \mathbb{Z}_2\times \mathbb{Z}_2$ instead of 
$\mathcal{D'} = \mathbb{Z}\times \mathbb{Z}$ because the N\'{e}el 
state \index{N\'{e}el antiferromagnet} transforms trivially under two-site
translations $(t_x)^2,(t_y)^2$. 
Thus, only the representations of $\mathcal{D'}$ trivial under
two-site translations are relevant; these are 
exactly the representations of $\mathcal{D}$. 
Put differently we only have to consider the
translations in the unitcell of the magnetic structure which in the
present case can be chosen as a $2$-by-$2$
cell. Furthermore, we will for now neglect pointgroup symmetries like
rotations and reflections of the lattice to simplify calculations.
At the end of this section we give results where also these symmetry elements 
are incorporated.

The groundstate manifold $V_{\text{GS}}$ 
we consider are the states related to $\ket{\psi}$ by an element of the symmetry
group $\mathcal{G}$, i.e.
\begin{equation}
  \label{eq:neelmanifold}
  V_{\text{GS}} = \left\{O_g\ket{\psi} ; g \in \mathcal{G}\right\}
\end{equation}
The symmetry elements in $\mathcal{G}$  that leave our prototypical state 
$\ket{\psi}$ invariant are given by two sets of elements:
\begin{itemize}
\item No translation in real space or a diagonal $t_{xy}$ translation
  together with a spin rotation $R_z(\alpha)$ around the $z$-axis 
  with an arbitrary angle $\alpha$.
\item Translation by one site, $t_x$ or $t_y$, followed by a rotation 
  $R_a(\pi)$ of $180^\circ$ around an axis $a \perp z$ perpendicular to the $z$-axis.
\end{itemize}
So the stabilizer of our prototype state $\ket{\psi}$ is given by
\begin{equation}
  \label{eq:neelstab}
  \text{Stab}(\ket{\psi}) = \left\{1\times R_z(\alpha)\right\}\cup \left\{t_{xy}\times R_z(\alpha)\right\}\cup\left\{t_x\times R_a(\pi)\right\}\cup\left\{t_y\times R_a(\pi)\right\}
\end{equation}
The representations of the discrete symmetry group can be 
labeled by four momenta $\mathbf{k} \in \left\{ (0,0) ,\; (0,\pi)  ,\; (\pi, 0)
  ,\; (\pi,\pi) \right\}$  with corresponding characters 
\begin{equation*}
  \chi_{\mathbf{k}}(t) = \text{e}^{i\mathbf{k}\cdot \mathbf{t}}
\end{equation*}
where $\mathbf{t}$ denotes the translation vector corresponding to $t$.
The continuous symmetry group we consider is the Lie group SO($3$). Its 
representations are labeled by the total spin $S$. The character of
the spin-$S$ representation is given by 
\begin{equation*}
  \chi_S(R) = \frac{\sin\left[(S+\frac{1}{2})\phi\right]}{\sin\frac{\phi}{2}}
\end{equation*}
where $\phi \in [0, 2\pi]$ is the angle of rotation of the element $R
\in \text{SO(}3\text)$. We see that spin rotations with different axes 
but same rotational angle give rise to the same character.
The representations of the total symmetry group $\mathcal{G} = \mathcal{D}
\times \mathcal{C}$ are now just the product representations of $
\mathcal{D}$ and $\mathcal{C}$. Therefore also the characters of
representations of $\mathcal{G}$ are the product of characters 
of $\mathcal{D}$ and $\mathcal{C}$. We label these representations
by $(\mathbf{k},S)$ where $\mathbf{k}$ denotes the lattice momentum and $S$ the
total spin.
To derive the multiplicities of the representations $(\mathbf{k},S)$ in the
groundstate manifold, we now apply the character-stabilizer formula,
Eq.~\eqref{eq:stabilizerformula}.
In the case of the square antiferromagnet this yields 
\begin{align}
  \label{eq:neelcharacter}
  n_{(\mathbf{k},S)}\, =\quad& e^{i\mathbf{k}\cdot 0}
	      \frac{1}{4\left| R_z(\alpha)\right|}
              \int\limits_0^{2\pi}d\alpha \chi_S(R_z(\alpha)) +
              e^{i\mathbf{k}\cdot (\mathbf{e}_x + \mathbf{e}_y)}
	      \frac{1}{4\left|R_z(\alpha) \right|}
              \int\limits_0^{2\pi}d\alpha \chi_S(R_z(\alpha)) \\
              &+  e^{i\mathbf{k}\cdot \mathbf{e}_x}
	      \frac{1}{4\left| R_a(\pi)\right|}
                 \int\limits_0^{2\pi}da \chi_S(R_a(\pi)) +
                 e^{i\mathbf{k}\cdot \mathbf{e}_y}
		 \frac{1}{4\left| R_a(\pi)\right|}
                 \int\limits_0^{2\pi}da \chi_S(R_a(\pi))
\end{align}

We compute 
\begin{equation*}
  \left| R_z(\alpha) \right| =  \left| R_a(\pi) \right|= \int\limits_0^{2\pi}d\phi = 2\pi
\end{equation*}
\begin{equation}
  \frac{1}{2\pi} \int\limits_0^{2\pi}d\alpha \chi_S(R_z(\alpha)) =
\frac{1}{2\pi}\int\limits_0^{2\pi}d\alpha
\frac{\sin\left[(S+\frac{1}{2})\alpha\right]}{\sin\frac{\alpha}{2}}  
  =  \frac{1}{2\pi}\int\limits_0^{2\pi}d\alpha \sum\limits_{l=-S}^S e^{il\alpha} = 1
 \label{eq:char_int1}
\end{equation}
and
\begin{equation}
  \frac{1}{2\pi} \int\limits_0^{2\pi}da \chi_S(R_a(\pi)) =
\frac{1}{2\pi}\int\limits_0^{2\pi}da
\frac{\sin\left[(S+\frac{1}{2})\pi\right]}{\sin\frac{\pi}{2}} = (-1)^S
 \label{eq:char_int2}
\end{equation}
Putting this together gives the final result for the multiplicities of
the representations in the tower of states
\begin{align}
  n_{\left((0,0),S\right)}& = \frac{1}{4}\left(1\cdot 1 + 1\cdot1 +
                            1\cdot (-1)^S + 1\cdot (-1)^S\right) =  
                            \left\{ \begin{array}{l} 1 \text{ if } S
                                      \text{ even} \\ 
                                      0 \text{ if } S \text{ odd}
                                    \end{array} \right. \\
  n_{\left ((\pi,\pi),S\right)} &= \frac{1}{4}\left(1\cdot 1 + 1\cdot1
                                  - 1\cdot (-1)^S - 1\cdot (-1)^S\right)  =
                                  \left\{ \begin{array}{l} 0 
                                            \text{ if } S 
                                            \text{ even} \\ 1 \text{ if } S \text{ odd}
                        \end{array} \right. \\
   n_{\left ((0,\pi),S\right)} &= \frac{1}{4}\left(1\cdot 1 - 1\cdot1
                                  + 1\cdot (-1)^S - 1\cdot
                                 (-1)^S\right) = 0\\
   n_{\left ((\pi,0),S\right)} &= \frac{1}{4}\left(1\cdot 1 - 1\cdot1
                                  - 1\cdot (-1)^S + 1\cdot
                                 (-1)^S\right) = 0
\end{align}
Tab.~\ref{tab:TOS_Neel} lists the computed multiplicities of the irreducible
representations where additionally the $\text{D}_4$ point group was considered
in the symmetry analysis. These irreps and their multiplicities exactly agree
with the irreps and multiplicities in the TOS of the square lattice
Heisenberg model from ED in Fig.~\ref{fig:TOS_Neel}. The spectroscopic
predictions together with the numerical data thus constitute a firm and solid
evidence of N\'{e}el order.


\begin{table}[ht]
 \centering
  \begin{tabular}{c|cc}  
  $S$ & $\Gamma$.A1 & $M$.A1\\ \hline
  0 & 1 & 0\\
  1 & 0 & 1\\
  2 & 1 & 0\\
  3 & 0 & 1\\
  \end{tabular}
 \caption{Multiplicities of irreducible representations in the TOS for the
N\'eel Antiferromagnet on a square lattice.}
 \label{tab:TOS_Neel}
\end{table}

\subsubsection{Magnetic order on the triangular lattice}\label{sec:magtri}
On the triangular lattice several magnetic orders can be
stabilized. The Heisenberg nearest neighbour model has been
shown to have a $120^\circ$ N\'{e}el \index{triangular antiferromagnet} 
ordered groundstate where 
spins on neighbouring sites align in an angle of  
$120^\circ$~\cite{Jolicoeur1990,Chubukov1992}.
Upon adding further second nearest neighbour interactions $J_2$ to
the Heisenberg nearest neighbour model with interaction strength $J_1$
it was shown that the groundstate exhibits \textit{stripy order} for
$J_2/J_1 \gtrsim 0.18$~\cite{Lecheminant1995}. 
Here spins are aligned ferromagnetically along
one direction of the triangular lattice and antiferromagnetically
along the other two. Interestingly, it was shown that a phase exists between
these two magnetic orders whose exact nature is unclear until today. Several
articles propose that in this region an exotic \textit{quantum spin
liquid} is stabilized \cite{Iqbal2016,Kaneko2014,Hu2015a,Zhu2015}. 
In a recent proposal two of the authors established an
approximate phase diagram of an extended Heisenberg model
with further scalar chirality interactions 
$J_{\chi}\mathbf{S}_i\cdot(\mathbf{S}_j\times \mathbf{S}_k)$~\cite{Wietek2016}
on elementary triangles. The Hamiltonian of this model is given
by 
\begin{equation}
    \label{eq:hamiltonianj1j2jch}
    H \,=\, J_1\sum\limits_{\left<
i,j\right>}\mathbf{S}_i\cdot\mathbf{S}_j \ + 
    J_2\sum\limits_{\left<\left<
    i,j\right>\right>}\mathbf{S}_i\cdot\mathbf{S}_j +
    J_{\chi}\sum\limits_{i,j,k \in \bigtriangleup}
\mathbf{S}_i\cdot(\mathbf{S}_j\times \mathbf{S}_k)
\end{equation}
Amongst the already known $120^\circ$ N\'{e}el and stripy phases
an exotic \textit{Chiral Spin Liquid} and a magnetic
\textit{tetrahedrally ordered} phase were found. \index{tetrahedral order}
Here we will only discuss the magnetic orders appearing in this model.
The non-coplanar tetrahedral order has a four-site unitcell where four spins align such 
that they span a regular tetrahedron. In this chapter we discuss the
tower of states for the three magnetic phases in this model. 
 
\begin{figure}[ht]
 \centering
 \begin{subfigure}[c]{.5\textwidth}
 \includegraphics[width=\textwidth]{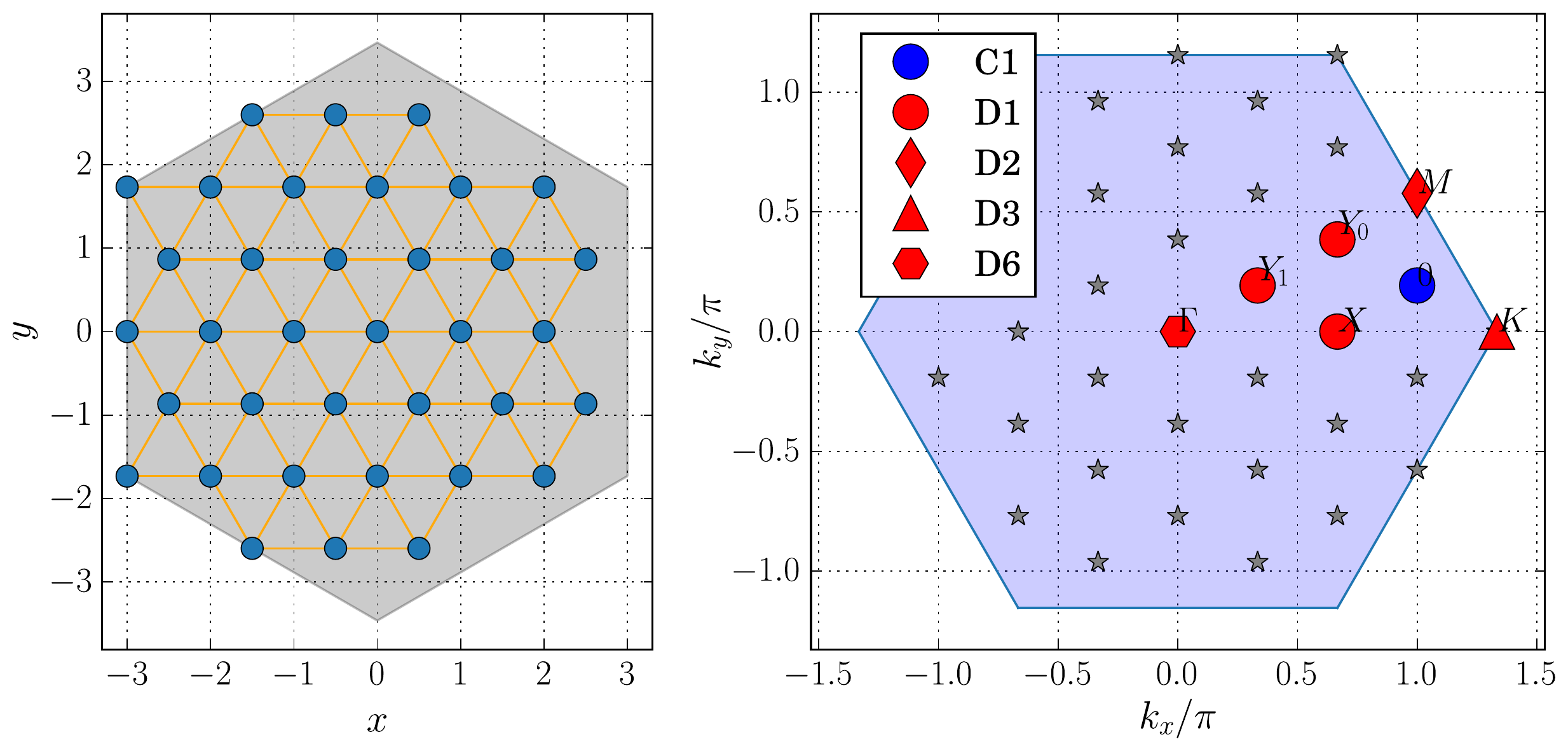} 
 \end{subfigure}
 \hfill
 \begin{subfigure}[c]{.45\textwidth}
 \includegraphics[width=\textwidth]{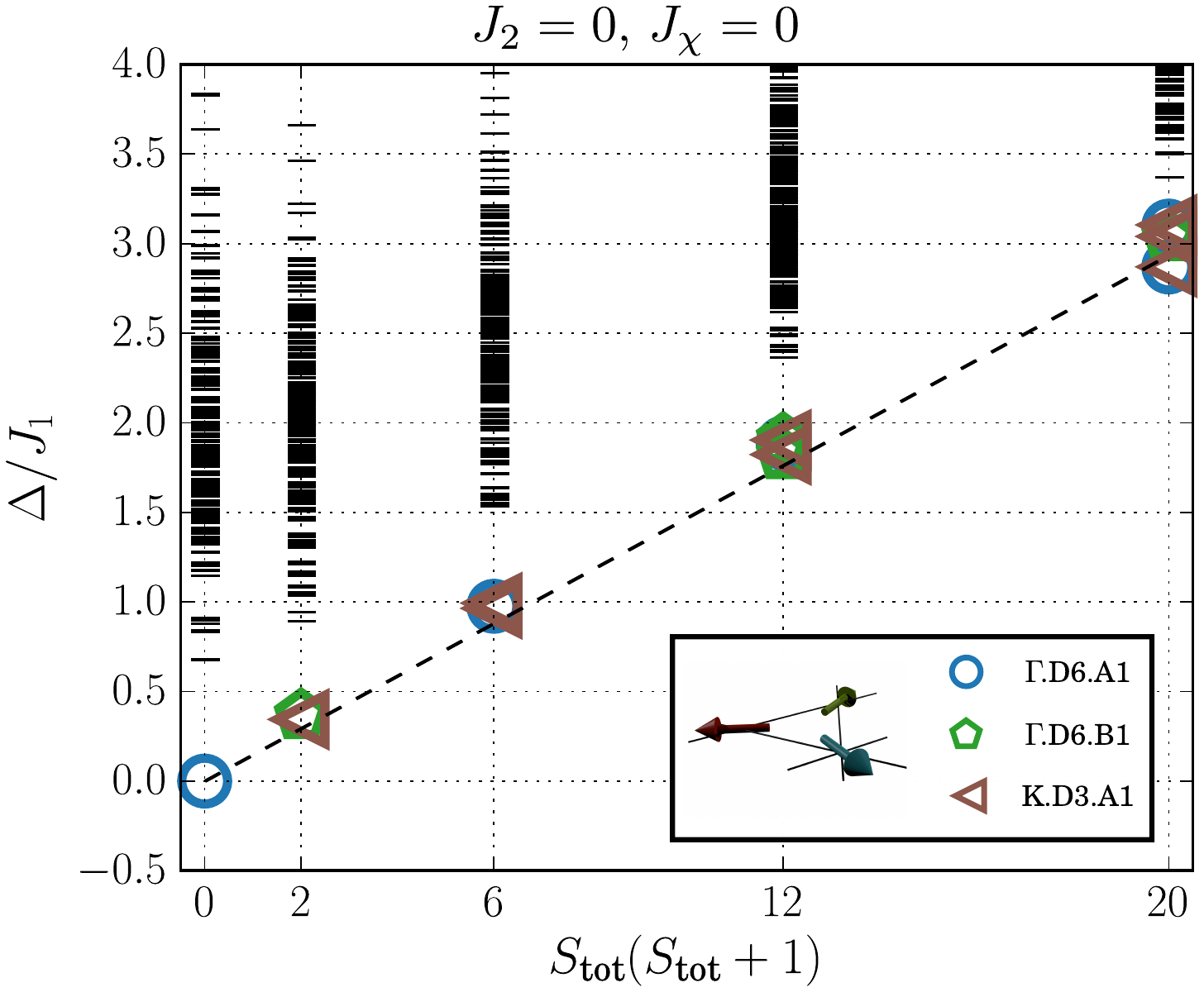}
 \end{subfigure}
 \caption{(Left): Simulation cluster for the Exact Diagonalization
calculations. (Center): Brillouin zone of the triangular lattice with the
momenta which can be resolved with this choice of the simulation cluster.
Different symbols denote the little groups of the corresponding momentum.
(Right): TOS for the $120^{\circ}$ N\'eel order on the triangular lattice. The
symmetry sectors and multiplicities fulfill the predictions from the
symmetry analysis (See Tab.~\ref{tab:andersontowermagtri}). 
One should note, that the
multiplicities grow with $S_{\text{tot}}$ for non-collinear states.}
 \label{fig:3sublatt_triangular}
\end{figure}

\begin{figure}[ht]
 \centering
 \begin{subfigure}[c]{.45\textwidth}
 \includegraphics[width=\textwidth]{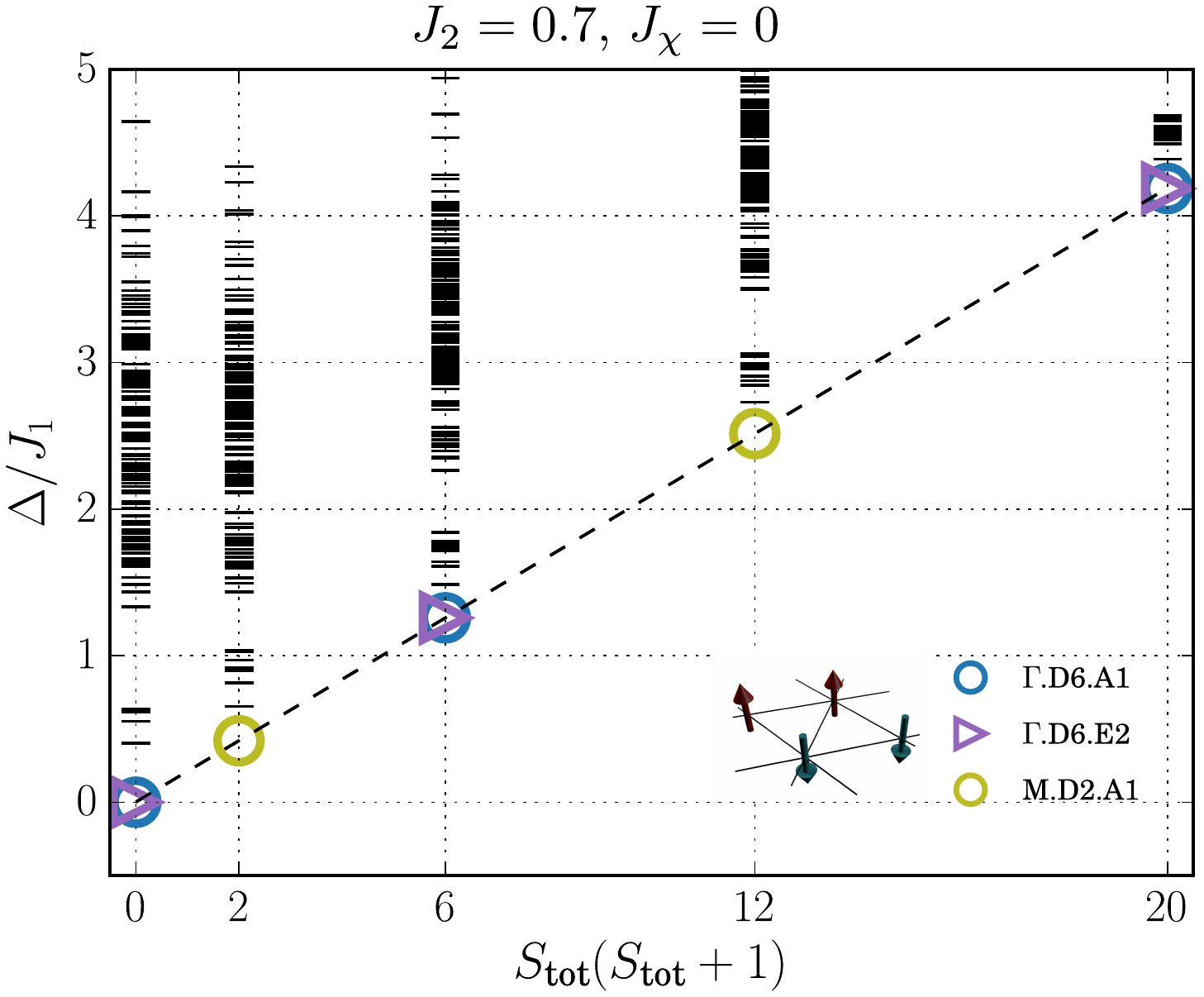}
 \end{subfigure}
 \begin{subfigure}[c]{.45\textwidth}
 \includegraphics[width=\textwidth]{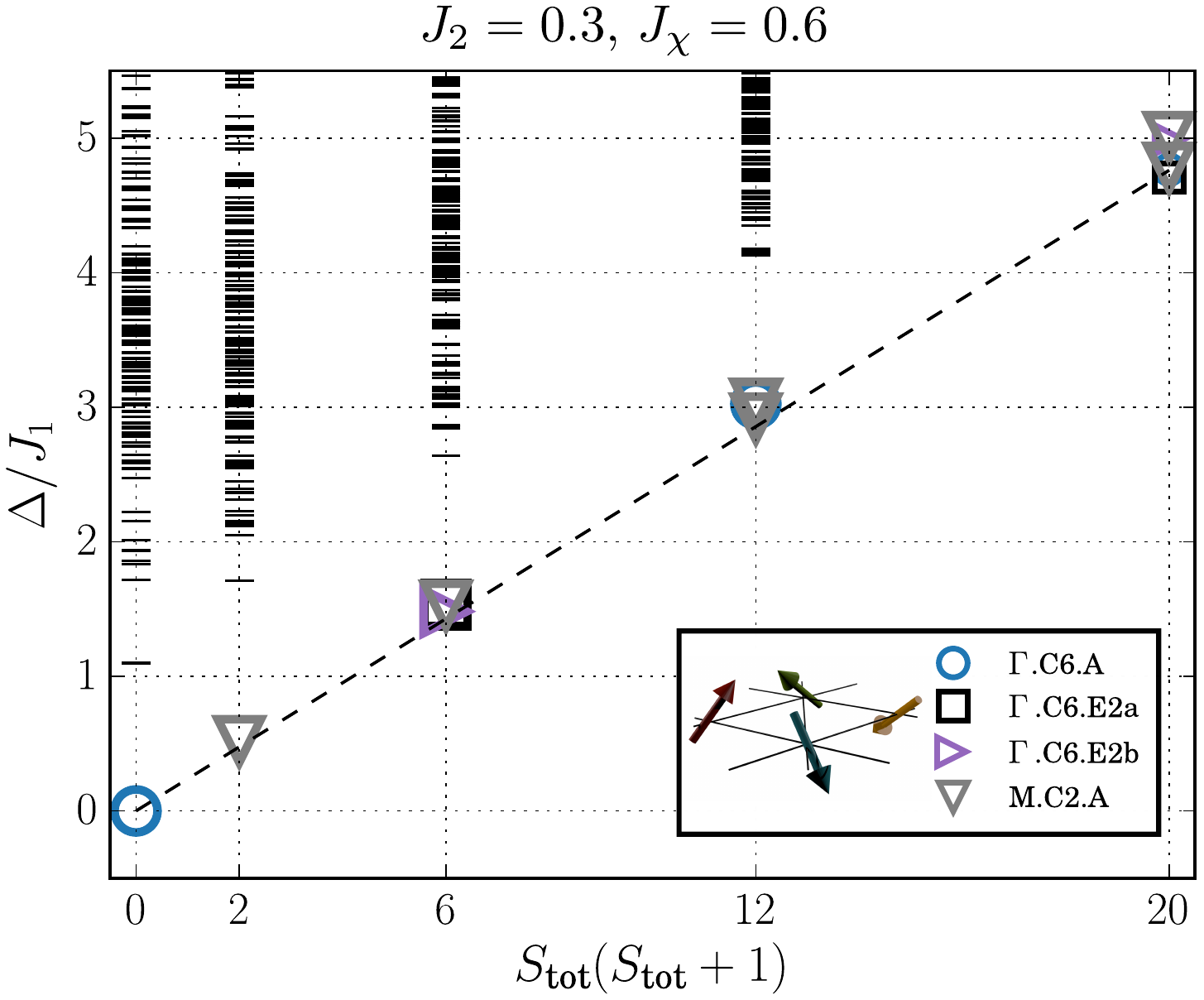}
 \end{subfigure}
 \caption{(Left): TOS for the stripy phase on the triangular lattice. The
multiplicities for each even/odd $S_{\text{tot}}$ are constant for collinear
phases. (Right): TOS for the tetrahedral order on the triangular
lattice.}
 \label{fig:stripy_tetra_triangular}
\end{figure}

First of all Fig.~\ref{fig:3sublatt_triangular} shows the simulation cluster
used for the Exact Diagonalization calculations
in \cite{Wietek2016}. We chose a $N=36=6\times 6$ sample with periodic boundary 
conditions. This sample allows to resolve the momenta $\Gamma$, $K$ 
and $M$, amongst several others in the Brillouin zone. The $K$ and $M$ 
momenta are the ordering vectors for the $120^\circ$, stripy and
tetrahedral order. Furthermore, this sample features full sixfold
rotational as well as reflection symmetries (the latter only in the absence of the chiral term, i.e. $J_{\chi} = 0$). Its pointgroup is 
therefore given by the dihedral group of order 12, D$_6$. 
The little groups of the individual $\mathbf{k}$ vectors are also shown in 
Fig.~\ref{fig:3sublatt_triangular}. 
For our tower of states analysis we now want to consider 
the discrete symmetry group 
\begin{equation}
        \mathcal{D} = \mathcal{T} \times \text{D}_6
\end{equation}
where $\mathcal{T}$ is the translational group of the magnetic
unitcell. The full set of irreducible representations of this 
symmetry group is given by the set $(\mathbf{k}\otimes\rho_{\mathbf{k}})$ where
$\mathbf{k}$ denotes the momentum and $\rho_{\mathbf{k}}$ is an irrep
of the little group associated to $\mathbf{k}$. The points $\Gamma$, $K$
and $M$ give rise to the little groups D$_6$, D$_3$ and D$_2$ 
(the dihedral groups of order $12$, $8$, and $4$), respectively.
For the stripy and tetrahedral order we can choose a 
$2\times2$ magnetic unitcell, and a $3\times3$ unitcell for 
the $120^\circ$ N\'{e}el order. The spin rotational symmetry 
gives rise to the continuous symmetry group
\begin{equation}
        \mathcal{C} = \text{SO(}3\text{)}
\end{equation}
We can therefore label the full set of irreps as $(\mathbf{k},
\rho_{\mathbf{k}}, S)$
where $S$ denotes the total spin $S$ representation of SO($3$).
Similarly to the previous chapter we now want to apply the 
character-stabilizer formula, Eq.~\eqref{eq:stabilizerformula},
to determine the multiplicities
of the representations forming the tower of states. The characters
of the irreps $(\mathbf{k},\rho_{\mathbf{k}},S)$ are given by 
\begin{equation}
        \chi_{(\mathbf{k},\rho_{\mathbf{k}},S)}(t, p, R) =
\text{e}^{i\mathbf{k}\cdot \mathbf{t}}
        \chi_{\rho_{\mathbf{k}}}(p)
        \frac{\sin\left[(S+\frac{1}{2})\phi\right]}{\sin\frac{\phi}{2}}
\end{equation}
where again $\phi \in [0, 2\pi]$ is the angle of rotation of the spin
rotation $R$. 
\begin{table}[htb]
  \begin{center}
    \begin{tabular}{c|c|c|c|c|c|c}
      $\text{D}_6$ & 1 & $2C_6$ & $2C_3$ & $C_2$ &
                                                              $3\sigma_d$ & $3\sigma_v$\\ \hline
      A1 & 1 & 1 & 1 & 1 & 1 & 1 \\
      A2 & 1 & 1 & 1 & 1 & -1 & -1 \\
      B1 & 1 & -1 & 1 & -1 & 1 & -1 \\
      B2 & 1 & -1 & 1 & -1 & -1 & 1 \\
      E1 & 2 & 1 & -1 & -2 & 0 & 0 \\
      E2 & 2 & -1 & -1 & 2 & 0 & 0 \\
    \end{tabular}
    \caption{Character table for pointgroup D$_6$.}
    \label{tab:d6chartable}
  \end{center}
\end{table}
The characters of the pointgroup D$_6$ are given in 
Tab.~\ref{tab:d6chartable}.
We skip the exact calculations which follow
closely the calculations performed in the previous chapter,
although now pointgroup symmetries are additionally
taken into account. The results are summarized in 
Tab.~\ref{tab:andersontowermagtri}.
\begin{table}[h]
  \centering
  \begin{tabular}{|l|*{3}{c}c|*{3}{c}c|*{4}{c}|}
    \hline
    &\multicolumn{4}{l|}{$120^\circ$ N\'{e}el} 
    &\multicolumn{4}{l|}{stripy order} 
    & \multicolumn{4}{l|}{tetrahedral order}\\
    \hline\hline
      $S$ & $\Gamma$.A1 & $\Gamma$.B1 & K.A1  & \quad 
              & $\Gamma$.A1 & $\Gamma$.E2 & M.A   & \quad
              & $\Gamma$.A & $\Gamma$.E2a & $\Gamma$.E2b & M.A  \\
    \hline
    0   &  1  & 0 & 0 & \quad & 1 & 1 & 0 & \quad & 1&0&0&0\\
    1   &  0  & 1 & 1 & \quad & 0 & 0 & 1 & \quad & 0&0&0&1\\
    2   &  1  & 0 & 2 & \quad & 1 & 1 & 0 & \quad & 0&1&1&1\\
    3   &  1  & 2 & 2 & \quad & 0 & 0 & 1 & \quad & 1&0&0&2\\
    \hline
  \end{tabular}
  \caption{Multiplicities of irreducible representations in the
    Anderson tower of states for the three magnetic orders on the triangular
    lattice defined in the main text. }
  \label{tab:andersontowermagtri}
\end{table}
We remark that the tetrahedral order is stabilized only for $J_\chi
\neq 0$ where the model in Eq.~\eqref{eq:hamiltonianj1j2jch}
does not have reflection symmetry any more since the term
$\mathbf{S}_i\cdot(\mathbf{S}_j\times \mathbf{S}_k)$ does not preserve this
symmetry. Therefore we used only the pointgroup C$_6$ of sixfold
rotation in the calculations of the tower of states for this phase. 

If we compare these results to
Figs.~\ref{fig:3sublatt_triangular},\ref{fig:stripy_tetra_triangular} we see
that these are exactly the representations appearing in the TOS from Exact 
Diagonalization for certain parameter values $J_2$ and $J_\chi$. This is a
strong evidence that indeed SO($3$) symmetry is broken in these models in a way
described by the $120^\circ$ N\'{e}el, stripy
and tetrahedral magnetic prototype states.

It is worth noting, that the sum of the multiplicities is constant with $S_{\text{tot}}$ for collinear phases, e.g. the stripy order shown here, whereas it is increasing for non-collinear orders.

\subsubsection{Quadrupolar order}\label{sec:quadru}
\index{quadrupolar order}
All examples of continuous symmetry breaking we have discussed so far
spontaneously broke $\text{SO}(3)$ symmetry but exhibited a magnetic moment. In
the
following we will show examples of phases that do not exhibit any magnetic
moment but break spin-rotational symmetry anyway and discuss the influences on
the tower of states. We will restrict our discussion to quadrupolar phases in $S=1$
models here, a broader introduction to nematic and multipolar phases can be
found in \cite{Penc2011}.

\paragraph{Quadrupolar states}
We denote the basis states for a single spin $S=1$ with $S^z = 1,-1,0$ as
$\ket{1}, \ket{\bar{1}}, \ket{0}$. In contrast to the usual $S=1/2$ case not
each basis state can be obtained by a $\text{SU}(2)$ rotation of any other basis
state. The state $\ket{0}$, for example cannot be obtained by a rotation of
$\ket{1}$ or $\ket{\bar{1}}$ as it has no orientation in spin-space at all,
$\braket{0|S^{\alpha}|0}=0$ \cite{Penc2011}. The state $\ket{0}$ can, however, 
be described as a spin fluctuating in the $x-y$ plane in spin space as 
\begin{equation}
 \braket{0|(S^x)^2|0} = \braket{0|(S^y)^2|0} = 1, \quad \braket{0|(S^z)^2|0}=0
\end{equation}
We can thus assign a director $\mathbf{d}$ along the $z$-axis to this state.
$\text{SU}(2)$ rotations change the director of such a state, but not its
property of being non-magnetic. These states are identified as quadrupolar
states as they can be detected by utilizing the quadrupolar
operator~\cite{Penc2011}
\begin{equation}
 Q^{\alpha \beta} = S^{\alpha}S^{\beta} + S^{\beta}S^{\alpha} -\frac{2}{3}
S(S+1)\delta_{\alpha \beta}
\end{equation}

To study the possible formation of an ordered quadrupolar phase on a lattice,
where the directors of the quadrupoles on each lattice site follow a regular
pattern, we consider the \textit{bilinear-biquadratic} model with Hamiltonian
\index{bilinear-biquadratic model}
\begin{equation}
 H = \sum_{\langle i,j \rangle} J \; \mathbf{S}_i \cdot \mathbf{S}_j + Q
\left(\mathbf{S}_i \cdot \mathbf{S}_j \right)^2
 \label{eq:bilin_biquad}
\end{equation}
and $S=1$. 
The second term in Eq.~\eqref{eq:bilin_biquad} can be rewritten in terms of the
elements of $Q^{\alpha \beta}$ which can be rearranged into a 5-component vector
$\mathbf{Q}$ such that
\begin{equation}
 \mathbf{Q}_i \cdot \mathbf{Q}_j = 2(\mathbf{S}_i \cdot \mathbf{S}_j)^2 +
\mathbf{S}_i \cdot \mathbf{S}_j - \frac{4}{3}
 \label{eq:Qprod}
\end{equation}
The expectation value of Eq.~\eqref{eq:Qprod} for
quadrupolar states on sites $i$ and $j$ is given in terms of their directors
$\mathbf{d}_{i,j}$ \cite{Penc2011}:
\begin{equation}
 \braket{\mathbf{Q}_i \cdot \mathbf{Q}_j} = 2 \left(\mathbf{d}_i \cdot
\mathbf{d}_j \right)^2 -\frac{2}{3}
\end{equation}
Therefore, the second term in Eq.~\eqref{eq:bilin_biquad} favours regular
patterns of the directors of quadrupoles. When such states are formed, they
spontaneously break $\text{SU}(2)$ symmetry without exhibiting any kind of
magnetic moment. The first term in Eq.~\eqref{eq:bilin_biquad}, on the other
hand, favours magnetic spin ordering as we have already discussed in previous
sections. 

The phase diagram of Eq.~\eqref{eq:bilin_biquad} on the triangular lattice shows
extended ferromagnetic, antiferromagnetic ($120^{\circ}$), ferroquadrupolar (FQ)
and antiferroquadrupolar (AFQ) ordered phases. 
\index{quadrupolar order!ferroquadrupolar} 
\index{quadrupolar order!antiferroquadrupolar}
In the FQ phase quadrupoles on each lattice site are formed with all directors
pointing in a single direction, whereas the directors form a $120^\circ$
structure in the AFQ phase. In the following, we will show that the FQ and AFQ
phases can be identified and distinguished from the spin ordered phases using
the TOS analysis on finite clusters.

\paragraph{TOS for quadrupolar phases}
The TOS for the FQ and AFQ phases can be expected to show similar behaviour as
the TOS for magnetically ordered states as both spontaneously break the
spin-rotational symmetry. 
If we identify the symmetry-broken quadrupolar phases with their directors
pointing in any direction in spin-space we can perform the symmetry analysis of
the TOS levels in a very similar manner as for spin-ordered systems in the
previous sections. There is, however, one important thing to consider: The
directors should not be considered to be described with vectors, but with axes;
a quadrupole is recovered (up to a phase) by rotations about an angle $\pi$
around any axis $a$ in the $xy$-plane:
\begin{equation}
 \text{e}^{i\pi S^a} \ket{0} = - \ket{0}
 \label{eq:rot_quadrupol}
\end{equation}
Thus, the stabilizer in Eq.~\eqref{eq:stabilizerformula} is different for
quadrupolar phases and the TOS shows a different structure.
This property makes it possible to distinguish, e.g., a magnetic $120^{\circ}$
phase from its quadrupolar counterpart, the AFQ phase using TOS analysis.

\begin{figure}[ht]
 \centering
 \begin{subfigure}[c]{.45\textwidth}
 \includegraphics[width=\textwidth]{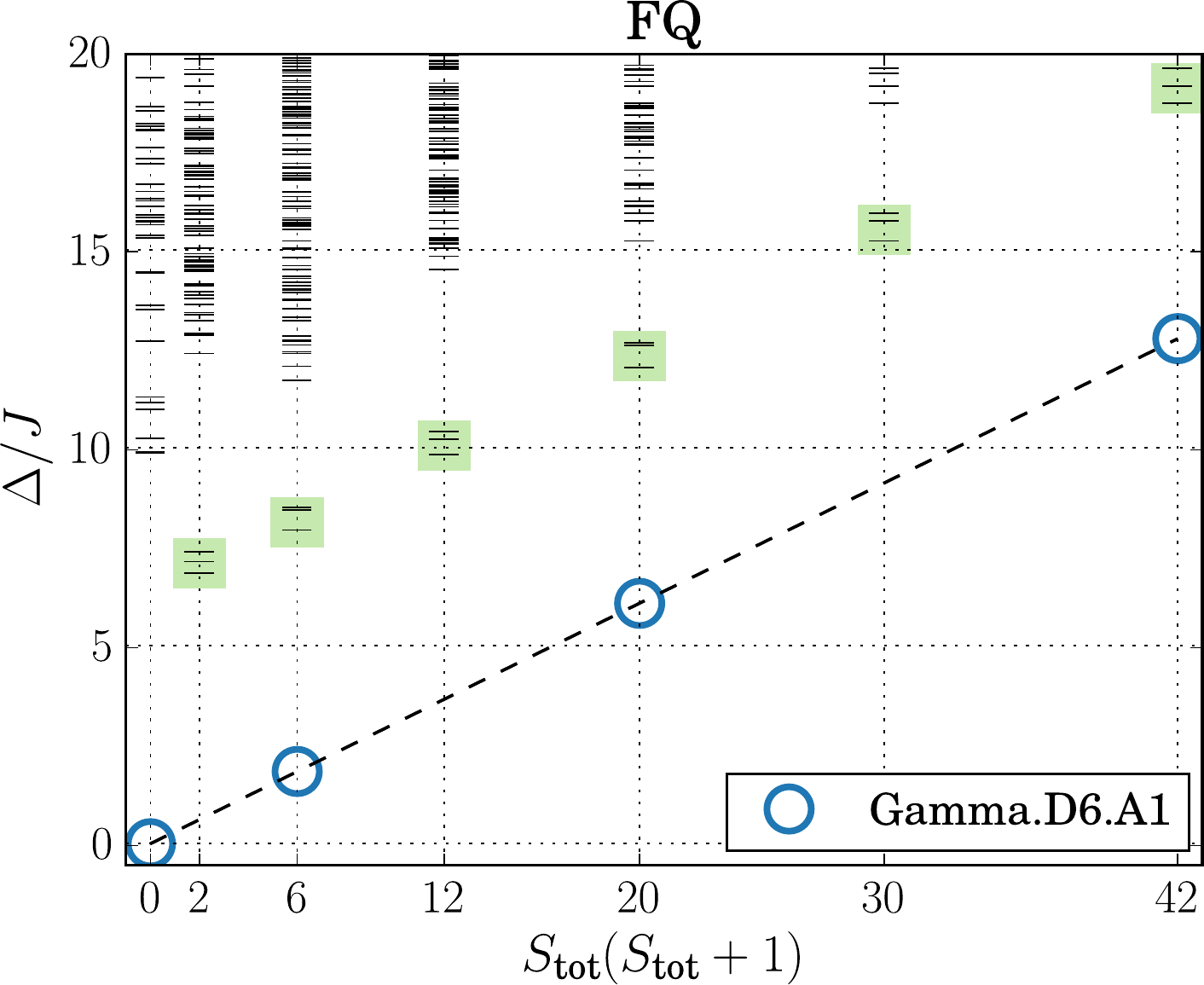}
 \end{subfigure}
 \hfill
 \begin{subfigure}[c]{.45\textwidth}
 \includegraphics[width=\textwidth]{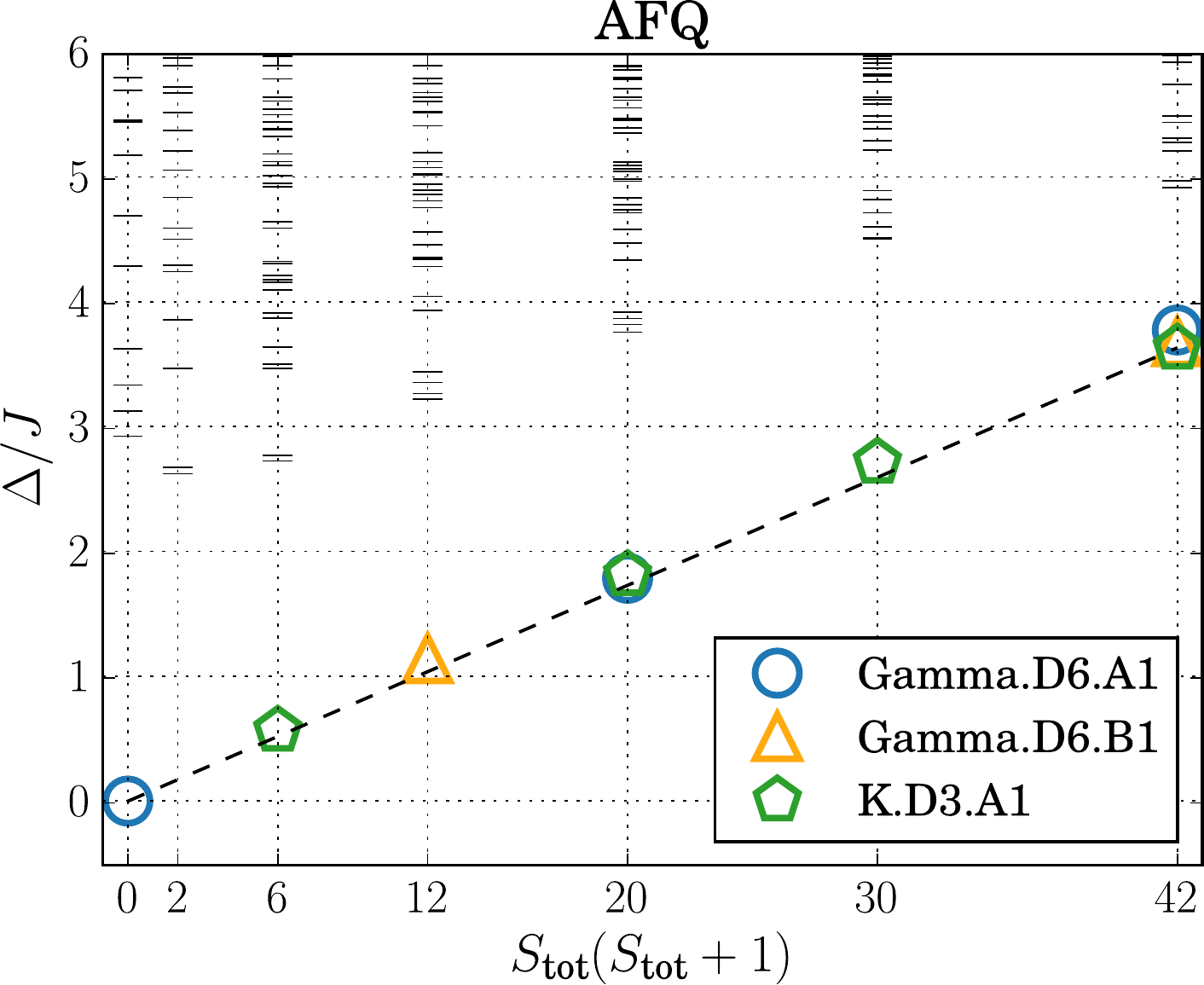}
 \end{subfigure}
 \caption{Tower of states for the ferroquadrupolar (left) and
antiferroquadrupolar (right) states on a triangular lattice with $N=12$
sites from Exact Diagonalization. The single-magnon branch for the FQ phase is
highlighted with green boxes.}
 \label{fig:TOS_quadrupolar}
\end{figure}

A prototype $\ket{\psi}$ for the FQ phase is a product states of quadrupoles
with directors in $z$-direction. This state does not break any space-group
symmetries, so only the trivial irreps of the space group,
$\mathbf{k}=\Gamma=(0,0), A1$, will be present in the TOS. The remaining
stabilizer of the spin-rotation group is a rotation around the $z$-axis about
an arbitrary angle and a rotation about an angle $\pi$ around any axis
lying in the $xy$-plane,
\begin{equation}
 \text{Stab}(\ket{\Psi}) = \{ R_z(\alpha), R_a(\pi) \}
\end{equation}
The multiplicities in the TOS can then be computed as 
\begin{align}
 n_{S} &= \frac{1}{2} \left( \frac{1}{|R_z(\alpha)|} \int_0^{2\pi} d\alpha
\chi_S(R_z(\alpha)) + (-1)^N \frac{1}{|R_a(\pi)|} \int_0^{2\pi} da
\chi_S(R_a(\pi)) \right) \\
&= \frac{1}{2} \left( 1 + (-1)^N(-1)^S \right)
\end{align}
where the integrals have already been computed in
Eqs.~\eqref{eq:char_int1} and \eqref{eq:char_int2}. The system size dependent
factor $(-1)^N$ is imposed from Eq.~\eqref{eq:rot_quadrupol}. To sum up, the TOS
for the FQ phase has single levels for even (odd) $S$ with trivial space-group
irreps and no levels for odd (even) $S$ sectors when $N$ is even
(odd)\footnote{For the simple case of the FQ phase one can also easily
calculate the decomposition of a state $\ket{S=1,m=0} \otimes \ket{S=1,m=0}
\otimes \dots$ into states $\ket{S_{tot},m=0}$ with the use of Clebsch-Gordan
coefficients.}.
The absence of odd (even) $S$ levels is caused by the invariance of quadrupoles
under $\pi$-rotation and distinguishes the TOS for a FQ phase from a usual
ferromagnetic phase. In Fig.~\ref{fig:TOS_quadrupolar} the computed TOS for the
model Eq.~\eqref{eq:bilin_biquad} in the FQ phase is shown on the left. It
shows the expected quantum numbers and multiplicities in the TOS and also an
easily identifiable magnon branch below the continuum.

The symmetry analysis for the AFQ phase can be performed in a similar manner
and shows a similar structure to the magnetic $120^{\circ}$-N\'eel phase, but
again levels are deleted for the AFQ. In this case, however, not all odd levels
are deleted but some levels in both, odd and even, $S$ sectors.
Tab.~\ref{tab:TOS_AFQ} shows the multiplicities of irreps in the TOS of the AFQ
model in comparison to the magnetic $120^{\circ}$-N\'eel state for even $N$.
Fig.~\ref{fig:TOS_quadrupolar} shows the simulated TOS for the AFQ phase for
the bilinear-biquadratic model Eq.~\eqref{eq:bilin_biquad}. The symmetry
sectors and multiplicities agree with the predictions.

\begin{table}[ht]
 \centering
 {%
\newcommand{\mc}[3]{\multicolumn{#1}{#2}{#3}}
\begin{tabular}{ccllcll}
 & \mc{3}{c}{AFQ} & \mc{3}{c}{$120^{\circ}$ N\'eel}\\\hline\hline
\mc{1}{c|}{S} & $\Gamma$.A1 & \mc{1}{c}{$\Gamma$.B1} & \mc{1}{c|}{$K$.A1} &
$\Gamma$.A1 & \mc{1}{c}{$\Gamma$.B1} & \mc{1}{c}{$K$.A1}\\\hline
\mc{1}{c|}{0} & 1 & \mc{1}{c}{0} & \mc{1}{c|}{0} & 1 & \mc{1}{c}{0} &
\mc{1}{c}{0}\\
\mc{1}{c|}{1} & 0 & \mc{1}{c}{0} & \mc{1}{c|}{0} & 0 & \mc{1}{c}{1} &
\mc{1}{c}{1}\\
\mc{1}{c|}{2} & 0 & \mc{1}{c}{0} & \mc{1}{c|}{1} & 1 & \mc{1}{c}{0} &
\mc{1}{c}{2}\\
\mc{1}{c|}{3} & 0 & \mc{1}{c}{1} & \mc{1}{c|}{0} & 1 & \mc{1}{c}{2} &
\mc{1}{c}{2}\\
 \end{tabular}
}
 \caption{Irreducible representations and multiplicities for the AFQ phase
compared to the magnetic $120^{\circ}$-N\'eel phase.}
 \label{tab:TOS_AFQ}
\end{table}

\section{Outlook}
In the previous sections we have discussed prominent features of the energy spectrum
for states which spontaneously break the spin-rotational symmetry in the
thermodynamic limit. We have seen that on finite-size systems the energy
spectra of such states exhibit a tower of states (TOS) structure. The tower of
states scales as $S_{\text{tot}}(S_{\text{tot}}+1)/N$ and generates the
groundstate manifold in the thermodynamic limit $N \rightarrow \infty$, which is
indispensible to spontaneously break a symmetry. The quantum numbers of the
levels in the TOS depend on the particular state which is formed after the
symmetry breaking and can be predicted using representation theory. 
\\[2ex]
As a generalization to the SU(2)-symmetric Heisenberg model,
Eq.~\eqref{eq:heisenberg}, one can introduce SU($n$) Heisenberg models with
$n>2$. Such models can experimentally be realized by ultracold multicomponent
fermions in an optical lattices. When the on-site repulsion is strong enough, the
Hamiltonian can be effectively described by an SU($n$) symmetric permutation
model on the lattice~\cite{Nataf2014}. If the exchange couplings are
antiferromagnetic, SU($n$) generalized versions of the N\'eel state might be
realized as groundstates, which then spontaneously break the SU($n$) symmetry of
the Hamiltonian. On finite systems this becomes again manifest in the emergence
of a tower of states, where the scaling is found to be proportional to
$C_2(n)/N$ \cite{Penc2003,Toth2010,Corboz2011,Corboz2013,Nataf2014}; $C_2(n)$
denotes the quadratic Casimir operator of SU($n$)\footnote{For $n=2$ the
quadratic Casimir operator $C_2 = S_{\text{tot}}(S_{\text{tot}}+1)$.}. The
symmetry analysis of the levels in the TOS can in principle be performed similar
to the case of SO(3) discussed in these notes but the symmetry group and its
characters have to be replaced with the more complicated group SU($n$).
\\[2ex]
On the other side, it can be also interesting to study models where the
continuous symmetry group is smaller. In real magnetic materials, the
isotropic Heisenberg interaction is often accompanied by other interactions
which, when they are strong enough, might reduce the symmetry group of spin
rotations to O(2); only spin rotations around an axis are a symmetry
of the system and can be spontaneously broken in the thermodynamic limit. This
symmetry group is also interesting in the field of ultracold gases, as BECs
spontaneously break an O(2) symmetry by choosing a phase. Tower of states can
also be found in this case and the quantum numbers and
multiplicities of the TOS levels can be computed in a similar fashion
\cite{Rousochatzakis2008}.
\\[2ex]

\begin{figure}[ht]
 \centering
 \includegraphics[width=.7\textwidth]
{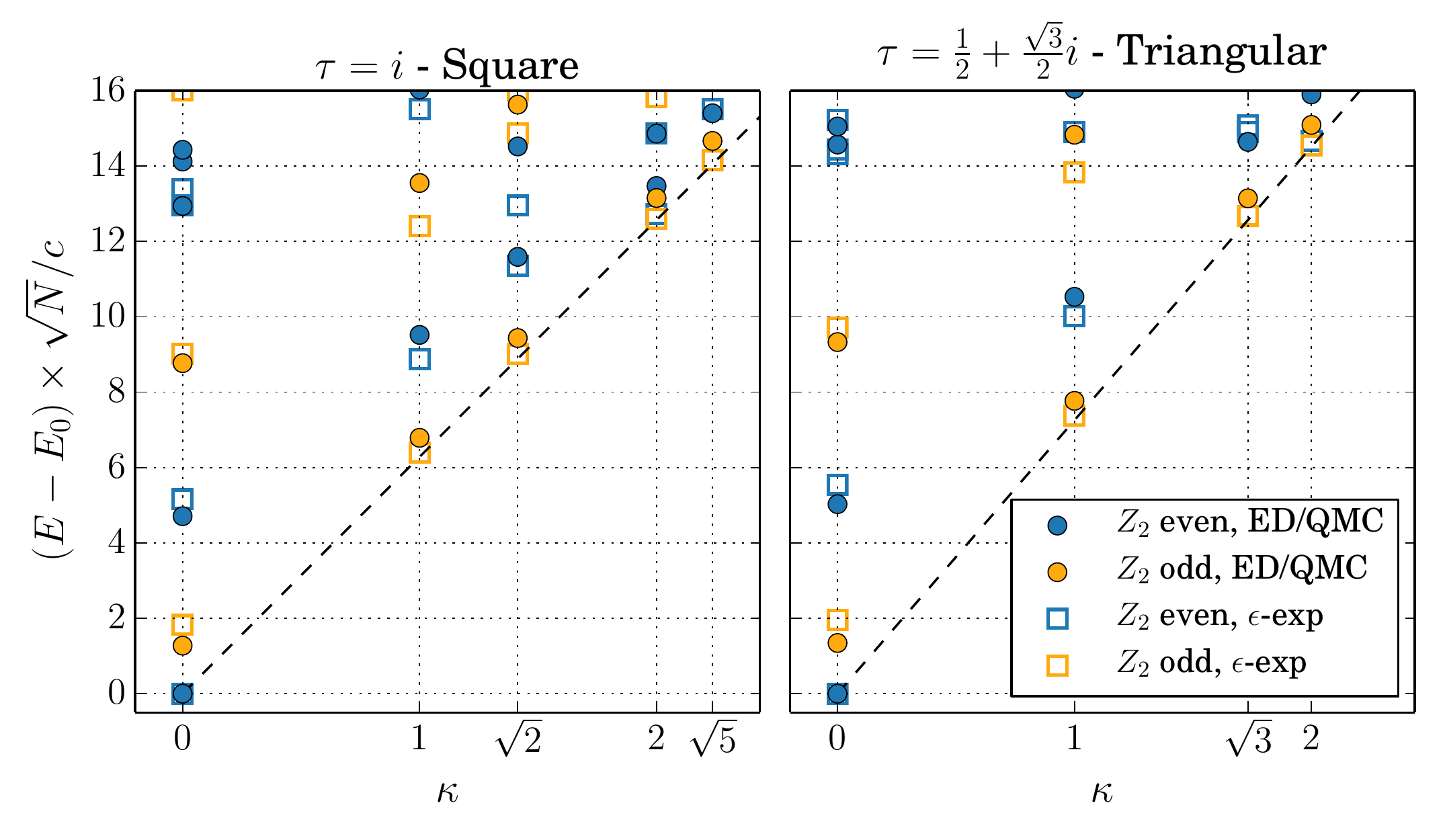}
 \caption{Universal torus spectrum for a continuous quantum phase transition in
the 3D Ising universality class. Full symbols denote numerical results while
empty symbols denote $\epsilon$-expansion results. The dashed line shows a
dispersion with the speed of light.}
 \label{fig:critical_ising}
\end{figure}

We have seen, that the energy spectrum of Hamiltonians on finite lattices
may contain a lot of information about the system. One can identify groundstates
which will spontaneously break discrete as well as continuous symmetries in the
thermodynamic limit and by imposing a classical state as symmetry broken state
one can even predict the quantum numbers and multiplicities of the levels in the
tower of states or in the quasi-degenerate groundstate manifold. 
When we impose an additional interaction to a system with spontaneously broken
groundstate, e.g. a magnetic field, it is possible that a
\textit{continuous quantum phase transition} (cQPT) \index{quantum phase
transition} from the ordered state to a
disordered state appears for some critical ratio of the couplings. Such cQPTs
are interesting as they can be described by universal features which do not
depend on most microscopic details of the model. Interestingly, the energy
spectrum on finite systems can even be used to identify and characterize cQPTs.
It is given by universal numbers times $1/L$, where $L=\sqrt{N}$ is the linear
size of the lattice. The quantum numbers of the energy levels show universal
features and are qualitatively related to the operator content of the
underlying \textit{critical field theory}, although the relation between them
is not yet fully understand for non-flat geometries, like a torus
\cite{Schuler2016, Whitsitt2016}. The critical spectrum for the transverse
field Ising model on a torus is shown in Fig.~\ref{fig:critical_ising}. It is a
fingerprint for the 3D Ising cQPT.

\clearpage
\bibliographystyle{correl}
\bibliography{lecture_notes_cont_symm_breaking}

\clearchapter


\end{document}